\documentclass[]{aa}  
\usepackage{natbib}
\bibpunct{(}{)}{;}{a}{}{,}
\usepackage{graphicx}
\usepackage{rotating}
\usepackage{txfonts}
\usepackage{multirow}
\usepackage[normalem]{ulem} 

\def\h2{H$_2~$2.12 $\mu$m}

\def\kms{km~s$^{-1}$}
\def\cm{cm$^{-3}$}
\def\cmsq{cm$^{-2}$}

\def\Lsun{L$_\odot$}
\def\Msun{M$_\odot$}
\def\Mw{$\dot{M}_{\rm w}$}
\def\yr{yr$^{-1}$}
\def\arcs{$^{\prime\prime}$}
\def\degree{\mbox{$^{\circ}$}}
\def\Vs {$V_{\rm s}$}

\def\Vexp {$V_{\rm exp}$}

\def\nH {$n_{\rm H}$}

\begin{document}
%
\title{Origin of the wide-angle hot H$_2$ in DG Tau}
\subtitle{New insight from SINFONI spectro-imaging} 
\author{
V. Agra-Amboage\inst{1}
          \and
          S. Cabrit\inst{2,4}
          \and
          C. Dougados\inst{3,4}
          \and
          L. E. Kristensen\inst{5}
          \and 
          L. Ibgui\inst{2}
          \and 
          J. Reunanen\inst{6}
}
\institute{Universidade do Porto, Faculdade de Engenharia, Departamento Engenharia fisica, 
SIM Unidade FCT 4006, Rua Dr. Roberto Frias, s/n 4200-465, Porto, Portugal
           \email{vaa@fe.up.pt}
           \and
             LERMA, UMR 8112 du CNRS \& Observatoire de Paris, ENS, UPMC, UCP, 
             61 Avenue de l'Observatoire, F-75014 Paris
             \and 
             Laboratoire Franco-Chilien d'Astronomie, UMI 3386 du CNRS, 1515 Camino el observatorio, Casilla 36-D correo central, Santiago, Chili
             \and
              IPAG, UMR 5274 du CNRS \& Univ. Joseph Fourier, 414 rue de la Piscine, F-38041 Grenoble
             \and 
             Harvard-Smithsonian Center for Astrophysics, 60 Garden Street, Cambridge, MA 02138, USA 
             \and 
             Tuorla Observatory, Department of Physics and Astronomy, 
             University of Turku, Vaisalantie 20, 21500 Piikkio, Finland 
}
\offprints{V. Agra-Amboage}
\date{Received; Accepted}

\abstract
{The origin of protostellar jets remains a major open question in star formation. Magneto-hydrodynamical (MHD) disk winds are an important mechanism to consider, because they would have a significant impact on planet formation and migration.} 
{We wish to test the origins proposed for the extended hot H$_{2}$ at 2000~K around the atomic jet from the  T Tauri star DG~Tau, in order to constrain the wide-angle wind structure and the possible presence of an MHD disk wind in this prototypical source.} 
{We present spectro-imaging observations of the DG~Tau jet in H$_{2}$ 1-0 S(1)  with 0\farcs12 angular resolution, obtained with SINFONI/VLT. Thanks to spatial deconvolution by the PSF and to careful correction for wavelength calibration and for uneven slit illumination (to within a few \kms), we performed a thorough analysis and modeled the morphology and kinematics. We also compared our results with studies in [Fe II], [O I], and  FUV-pumped H$_2$. Absolute flux calibration yields the H$_2$ column/volume density and emission surface, and narrows down possible shock conditions.}
{The limb-brightened H$_{2}$ 1-0 S(1) emission in the blue lobe is strikingly similar to FUV-pumped H$_2$ imaged 6 yr later, confirming that they trace the same hot gas and setting an upper limit $<12$\kms\ on any expansion proper motion. The wide-angle rims are at lower blueshifts (between -5 and 0 \kms) than probed by narrow long-slit spectra. We confirm that they extend to larger angle and to lower speed the onion-like velocity structure observed in optical atomic lines. The latter is shown to be steady over $\ge 4$ yr but undetected in [Fe II] by SINFONI, probably due to strong iron depletion.  The rim thickness $\le 14$ AU rules out excitation by C-type shocks, and J-type shock speeds are constrained to $\simeq$ 10 \kms.}
{We find that explaining the H$_2$ 1-0 S(1) wide-angle emission with a shocked layer requires either a recent outburst (15 yr) into a pre-existing ambient outflow or an excessive wind mass flux. A slow photoevaporative wind from the dense irradiated disk surface and an MHD disk wind heated by ambipolar diffusion seem to be more promising and need to be modeled in more detail. Better observational constraints on proper motion and rim thickness would also be crucial for clarifying the origin of this structure.}
\keywords{ star formation --
              protostellar outflows -- 
              molecular outflows --
              stars: individual: DG Tau --
              near-infrared spectroscopy --
              spectro-imaging -- 
               }
\maketitle
%

\section{Introduction}

 Jets from young stars are ubiquitous in star formation, and their link with accretion processes is well established \citep{Cabrit1990,Hartigan1995}. They are launched from the inner regions of the star-disk system, the same disk that may ultimately give rise to a planetary system. Although a number of models exist for their generation (see, for example, the reviews of \citet{Ferreira2006}, \citet{Shang2007} and \citet{Pudritz2007}), the exact mechanism by which mass is ejected from accreting systems and collimated into jets is an open problem. It is thought that the same basic magnetohydrodynamic (MHD) mechanism is responsible for  launching jets in objects as diverse as brown dwarfs \citep{Whelan2005}, post-AGB stars (e.g., \citet{Garcia-Diaz2008}, \citet{Garcia-Segura2005}), compact objects in binary systems, and perhaps even active galactic nuclei. 
Because of their proximity and rich set of diagnostic emission lines, jets from young stellar objects represent a particularly useful test bed for the MHD jet-launching paradigm \citep[see the reviews of][]{Bally2007,Ray2007}.

Molecular hydrogen is the primary gas constituent in the circumstellar environment of young stars.  The $v$=1-0 S(1) transition of  H$_2$ at 2.12 $\mu$m  is often detected in embedded protostars, with the emission formed primarily in shock-excited collimated outflows \citep[e.g.,][]{Davis2011,Davis2002}. The presence of H$_2$ in jets might be an indication that MHD ejection operates out to disk radii $\ge 0.2-1$AU where molecules can survive against dissociation \citep{Safier1993,Panoglou2012}. If such extended MHD disk winds persist until the later planet building phase, they could maintain fast accretion across the disk ``dead-zone" where ionization in the midplane is too low to activate MHD turbulence \citep{Bai2013b}. The associated magnetic fields would have strong consequences for the  formation and migration of planets \citep{Fromang2005,Bai2013}. The H$_2$ emission from warm (T=1000-3000 K) gas was recently detected within 100~AU of more evolved T Tauri stars (TTSs) and Herbig Ae/Be stars, both in the near-infrared and FUV domains \citep[e.g.,][]{Valenti2000, Ardila2002, Bary2003, Carmona2011}. When studied at high spectral resolution, these lines split into two populations: most show narrow profiles centered on the stellar velocity, suggesting an origin in a warm disk atmosphere (quiescent H$_2$), but $\simeq$ 30 \% of them display profiles blueshifted by 10-30 \kms, sometimes with blue wings extending up to -100 \kms, suggesting formation in an outflow \citep{Takami2004, Herczeg2006, Greene2010, Carmona2011, France2012}. In particular, long-slit spectra of DG Tau provide evidence of a warm H$_2$ outflow at $\simeq$ -20 \kms, shifted by $\simeq$ 30 AU along the blue jet direction and with an estimated width $\simeq$ 80 AU \citep{Takami2004, Schneider2013a}. 
\citet{Beck2008} conducted a 2D spectro-imaging study of H$_2$ rovibrational lines in six CTTs known to be associated with large scale atomic jets, including DG Tau, and found spatially extended emission in all cases, with sizes up to 200 AU. While the data are not flux calibrated, relative level populations among K-band H$_2$ lines indicate excitation temperatures in the range 1800 -- 2300 K. From various arguments, \citet{Takami2004} and \citet{Beck2008} argue that the H$_2$ properties are in general not well explained by FUV or Xray irradiation and are most consistent with shock-excited emission from the inner regions of the atomic jets or from wider-angle winds encompassing these flows. \citet{Takami2004} also mention ambipolar diffusion on the outer molecular streamlines of an MHD disk wind as an alternative heating mechanism for the warm H$_2$ outflow in DG Tau. 
\citet{Schneider2013b} also favor a molecular MHD disk wind, but heated by shocks rather than by ambipolar diffusion. In HL Tau, on the other hand, \citet{Takami2007} interpret the V-shaped H$_2$ emission encompassing the atomic jet as a shocked cavity driven into the envelope by an unseen wide-angle wind. In any case,  extended H$_2$ emission in TTS clearly holds key information on their wide-angle wind structure, the radial extension of any MHD disk wind present, and its impact on protoplanetary disk physics. 

    
In an effort to extract this information, we present in this paper an in-depth analysis of  H$_2$ ro-vibrational 1-0 S(1) emission associated with DG~Tauri, using flux-calibrated spectro-imaging K-band observations with 0\farcs12 resolution conducted with SINFONI/VLT. This work complements our detailed study of the atomic flow component and accretion rate in DG Tau based on [Fe II] lines in H-band and Br${\gamma}$ in K-band, observed at the same epoch with the same instrument \citep[][hereafter, Paper~I]{Agra-Amboage2011}.  Our analysis improves on that in \citet{Beck2008} in several ways: We apply spatial deconvolution to obtain a sharper view of the brightness distribution, and perform careful correction for wavelength calibration and uneven-slit illumination to retrieve radial velocities along and across the jet axis down to a precision of a few \kms. This allows us to conduct a detailed comparison with the spatio-kinematic structure of the atomic flow component and FUV H$_2$ emission in DG Tau, and to constrain both the proper motion and the 3D velocity field of warm H$_2$, including rotation. In addition, the absolute flux-calibration of the 1-0 S(1) line is used to infer column densities and emitting surfaces and, together with its ratio to  2-1 S(1)  at 2.25 $\mu$m, to restrict the possible shock parameters for H$_2$ excitation.  We then combine this new information to critically re-examine the main scenarii that have been invoked for the origin of wide-angle  H$_2$ emission in DG Tau and other TTS, namely: (1) an irradiated disk/envelope, (2) a molecular MHD disk wind heated by ambipolar diffusion, (3) a forward shock sweeping up the envelope, (4) a reverse shock driven back into a wide-angle molecular wind.

The paper is organized as follows: The observations, data reduction procedure, and centroid velocity measurements are described  in \S 2.  In \S 3 , we present the new results brought by our observations 
on the morphology, kinematics, and column density of the wide-angle H$_2$, and compare with the atomic  and FUV H$_2$ emission properties. In \S 4, we clarify the shock parameters that could explain the H$_2$ surface brightness and line ratio, we explore the possible 3D velocity fields, and we discuss the strengths and weaknesses of various proposed origins for the wide-angle H$_2$. Our conclusions are summarized in \S 5. Throughout the paper we assume a distance of 140 pc to the object. 


\section{H$_2$ observations, data reduction, and velocity measurements}\label{sec:obs}
\label{sec:reduc}
Observations of the DG~Tau microjet in K band were conducted on October 15th 2005 at the Very Large Telescope (VLT), using the integral field spectrograph SINFONI combined with an adaptive optics (AO) module \citep{sinfoni1,sinfoni2}. The field of view is 3\arcsec$\times$3\arcsec~with a spatial sampling of 0\farcs05$\times$0\farcs1 per "spaxel". The smallest sampling is along the direction of the slicing mirrors (denoted as X-axis in the following) and is aligned perpendicular to the DG Tau jet axis, i.e., at PA = 45\degr. After AO correction, the effective spatial resolution achieved in the core of the point spread function (PSF) is $\sim$0\farcs12 (FWHM of a Gaussian fit to the reconstructed continuum image).  The spectral configuration used provides a dispersion of 0.25nm/pixel over a spectral range  of 1.95 $\mu$m to 2.45 $\mu$m. 
The spectral resolution as determined from OH sky lines is $\sim$ 88 \kms\ (median FWHM), with smooth systematic variations of $\pm$ 15 \kms\ along the Y-axis. Three datacubes of 250s each were obtained, one on sky and two on source (referred to as exposure 1 and exposure 2 hereafter). 

The standard data reduction steps were carried out using the SINFONI pipeline. Datacubes are corrected for bad pixels, dark, flat field, geometric distortions on the detector, and sky background. An absolute wavelength calibration based on daytime arclamp exposures is also performed by the SINFONI pipeline. 
However, the accuracy is insufficient for kinematical studies of stellar jets.  Unlike for the H-band data of Paper I, the OH sky lines in our K-band sky exposure have too low signal-to-noise ratio (S/N) to provide a reliable improvement. Instead, we rely here on absorption lines of telluric and photospheric origin against the strong DG Tau continuum.  Telluric lines  near the H$_{2}$ 1-0 S(1) line were identified from comparison with a simulated atmospheric absorption spectrum convolved to our spectral resolution, generated with an optimized line-by-line code tested against laboratory measurements \citep{Ibgui_and_Hartmann_2002_a, Ibgui_and_Hartmann_2002_b}, using molecular data from the HITRAN database \citep{Rothman_et_al_2009}. Two unblended H$_2$O lines strong enough for velocity calibration were identified : one at $\simeq$ 2.112 $\mu$m (4735.229 cm$^{-1}$),\,and one at $\simeq$ 2.128 $\mu$m (4699.750 cm$^{-1}$). Three photospheric lines are also present on the blue side of the H$_{2}$ 1-0 S(1) line \citep{Photospheric}: a Mg doublet at $\simeq$ 2.107 $\mu$m (4746.841 and 4747.097 cm$^{-1}$) and two Al lines at $\simeq$ 2.110 $\mu$m and $\simeq$ 2.117 $\mu$m (4739.598 and 4723.769 cm$^{-1}$). All three are well detected in the DG Tau spectrum \citep[see also][]{Fischer2011}, but since the Mg line is a doublet we only use the two Al lines for velocity calibration. In each exposure, the centroids of the two Al photospheric lines and two H$_2$O telluric lines were determined by Gaussian fitting on a high S/N reference spectrum obtained by averaging spaxels over $\pm 0\farcs5$ along the slicing mirror closest to the DG Tau continuum peak (centered at $Y=0$ in exposure 1 and $Y=0\farcs03$ in exposure 2). After correcting photospheric lines for the motion of DG Tau \citep[$V_{\rm hel}$= 16.5 \kms,][]{Bacciotti2000}, the four velocity shifts were fitted by a linear function of wavelength to derive the global calibration correction to apply at 2.12 $\mu$m. The maximum deviation between the 4 data points and the linear fit is $\le 2$ \kms\ in each exposure, and we take this as an estimate of the maximum error in our absolute calibration.

The H$_2$ 1-0 S(1) emission profiles in each exposure are then continuum-subtracted and corrected for telluric absorption following the procedure described in Paper I: We select a high S/N reference "photospheric" spectrum with no discernable H$_2$ line at our resolution (at $\Delta x$$\sim$0\arcs, $\Delta y$$\sim$0\arcs). This reference spectrum is scaled down to fit the local continuum level of each individual spaxel, and subtracted out. This ensures optimal accuracy in the removal of telluric and photospheric absorption features (cf. Fig.~1 in Paper I); it will also remove the PSF of any faint H$_2$ emission present in the reference position, but the extended emission beyond 0\farcs1 of the star will not be affected. The same procedure is applied to extract the 2-1 S(1) line profiles. Correction for atmospheric differential diffraction is then done by registering each H$_2$ datacube on the centroid position of the nearby continuum, as calculated by a Gaussian fit in the $X$ and $Y$ directions of the reconstructed continuum image. The data are flux-calibrated on the standard star HD 28107, with an estimated error of 5\%. 

 H$_2$ centroid velocities are determined by a Gaussian fit to the extracted line profile. The fitted centroids are subject to a random fitting error due to noise, given by  $1\sigma \simeq$  FWHM / (2.35\,S/N) for a Gaussian of width FWHM and signal-to-noise ratio S/N  at the line peak \citep{Porter2004}. Monte-Carlo simulations show that this error can be much smaller than the velocity sampling \citep{Agra-Amboage2009}.
The FWHM of fitted Gaussians show the same systematic variations with Y-offset as OH sky lines and indicate a small intrinsic width $\Delta V  \leq$ 30 \kms\ for the H$_2$ emission, in agreement with higher resolution spectra \citep{Takami2004,Schneider2013a}.

After global absolute wavelength calibration, velocity centroids are still subject to small residual instrumental calibration drifts along/across slicing mirrors. Our simultaneous sky exposure has too low S/N to correct for this effect. We thus estimate its typical magnitude using bright OH lines near 2.12$\mu$m in two high S/N K-band sky datacubes obtained in the same instrumental configuration, during a different run. Along each slicing mirror, i.e., transverse to the jet axis, the centroids shifts are mainly caused by a systematic residual slope, with random fluctuations $< 1$\kms\ on top of it. After averaging along each mirror (i.e., when building a PV diagram along the jet axis), the OH centroids show random drifts from row ro row with an rms of 0.5--0.8 \kms. The resulting overall distribution of OH line centroids among individual spaxels in the field of view is essentially Gaussian, with an rms of 1.5 \kms. We will include the relevant random dispersion in our error bars on H$_2$ centroid velocities, and also take into account the systematic "slope effect" along slicing mirrors when searching for transverse rotation signatures (see Sect.~\ref{sec:vshift}). 

Finally, 
the H$_2$ velocity centroids need to be corrected for spurious velocity shifts due to {\em uneven-slit illumination}. Indeed, the slicing mirrors in SINFONI behave like a long-slit spectrograph where off-axis light suffers from a shift in wavelength on the detector with respect to on-axis light. The magnitude of this effect depends on the gradient of light distribution within the "slitlet", and in principle can be as high as the spectral resolution. In our observations, the total light distribution is dominated by the PSF of the stellar continuum, whose FWHM is on the order of the "slitlet" width (0.1$^{\prime\prime}$) therefore one might expect this effect to be significant. However, the AO system does not provide full correction for atmospheric seeing on DG Tau, so that most of the flux in the PSF is not in the Gaussian core but in broader wings with smoother gradients. As we demonstrate in Appendix~\ref{app:uesi}, the spurious velocity shifts in such conditions remain small (a few \kms). Furthermore, they can be modeled and corrected for with good accuracy, with an estimated residual error of $\leq$ 2 \kms, and even better in the blue lobe beyond $-0\farcs3$ from the star. All velocities in this paper are corrected for uneven slit illumination using the model outlined in Appendix~\ref{app:uesi}, and expressed in the reference frame of the star assuming $V_{\rm hel}$ = 16.5 \kms\ \citep{Bacciotti2000}. 

\section{Results}
\label{sec:results}

\subsection{H$_2$ 1-0 S(1) morphology and comparison with FUV and [Fe II] data} 
\label{sec:map}

\begin{figure*}
\centering 
\includegraphics[width=0.5\textwidth,trim=0 0 9.5cm 0,clip=true]{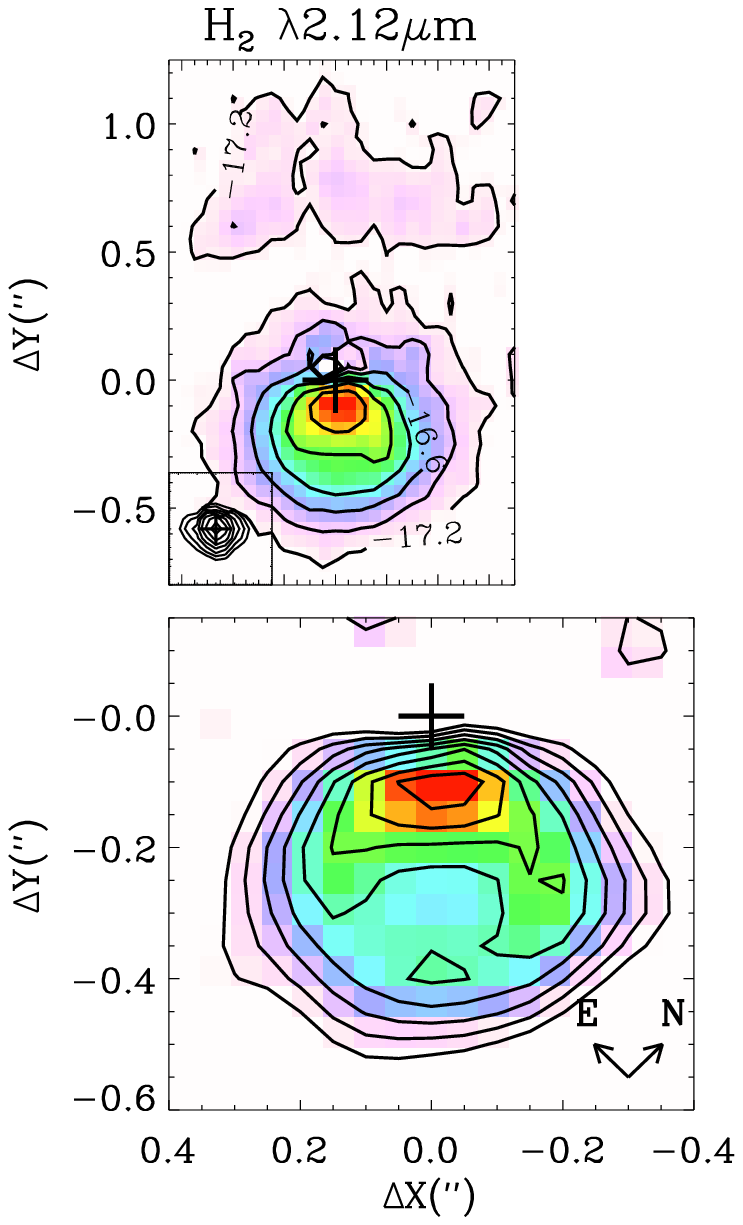}
\includegraphics[width=0.4\textwidth]{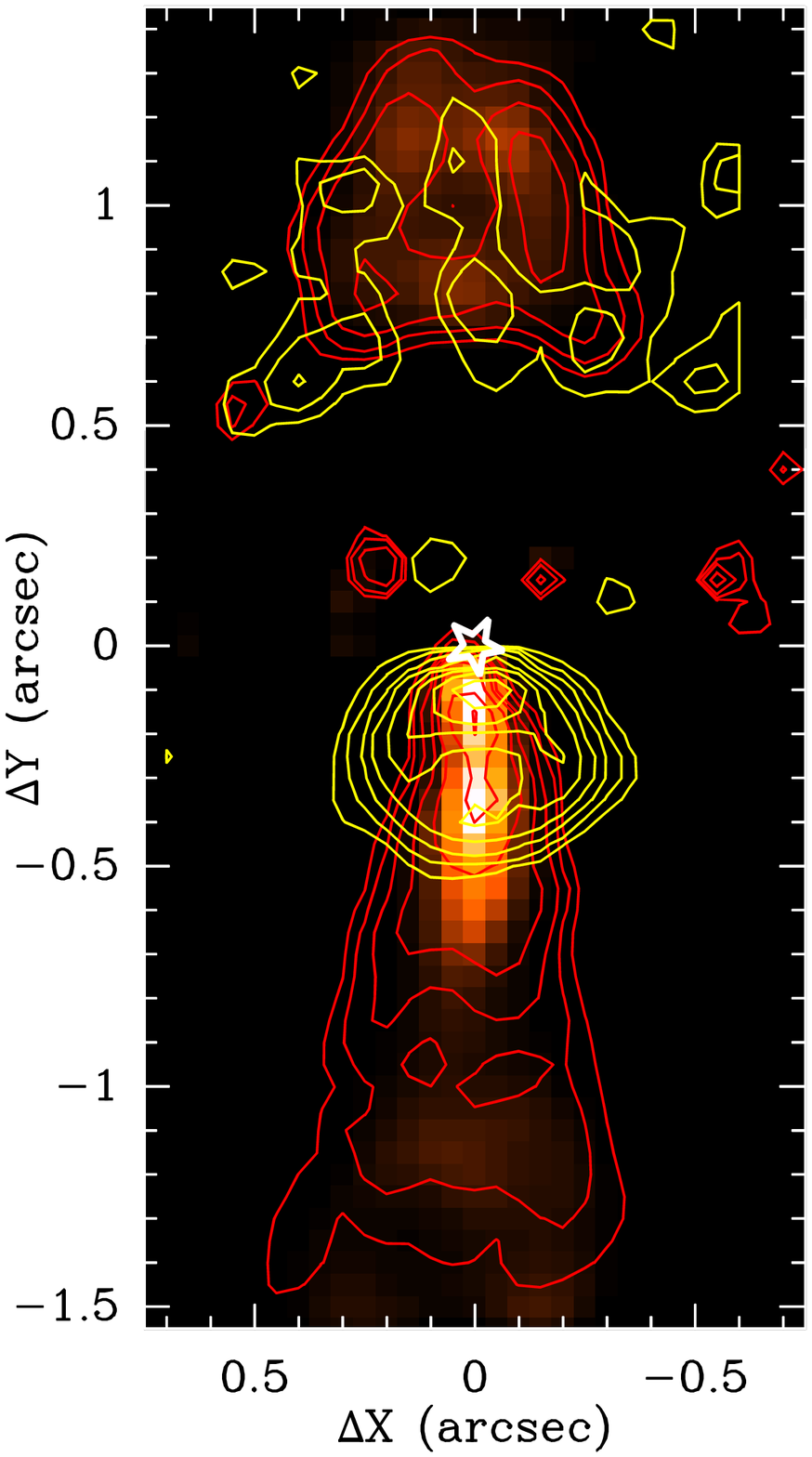}
\caption{Top left: Raw continuum-subtracted map of H$_{2}$ 1-0 S(1) 
 line emission in DG Tau. Contours start at 
 $2.7 \times 10^{-4}$ erg s$^{-1}$ cm$^{-2}$ sr$^{-1}$ 
and increase by factors of 2. The cross denotes the centroid of the continuum image (shown as an insert). Bottom left: H$_{2}$ 1-0 S(1) 
 map after deconvolution by the continuum image. The lack of H$_2$ emission at the central position results from our continuum subtraction procedure. Right panel: Deconvolved H$_{2}$ 1-0 S(1) 
image (yellow contours) superposed on the deconvolved channel maps of  [Fe~II]1.64 $\mu$m obtained at the same epoch and resolution by \citet{Agra-Amboage2011}, in two velocity ranges: MV (red contours; -160 \kms\ $< V <$ 120 \kms) and HV (background colour image;  $V < -160$ \kms\  and $V > 120$ \kms).}
\label{fig:channelsmap}
\end{figure*}

The top left panel of Fig.~\ref{fig:channelsmap} shows the raw continuum-subtracted map in the  H$_{2}$ 1-0 S(1) 
 line\footnote{Because the spectral response of SINFONI is non-Gaussian and includes broad  "shoulders" that may lie below the noise level, we integrated the line flux only over the three brightest velocity channels at -35.61, -0.61 and +34.38 \kms. A factor of 1.4 was then applied in order to account for the flux lost in these spectral wings (as estimated on the brightest positions in our map)}
where the two on-source exposures were combined to increase S/N. The morphology agrees with that first reported by \cite{Beck2008} in the same line at comparable resolution, namely: a bright cusp on the blue side (negative $\Delta Y$), a fainter broader arc on the red side at +0\farcs6--1\arcsec\ from the star, and a dark lane in between, presumably due to occultation by the circumstellar disk of DG Tau. The dark lane was previously seen in optical and H-band observations of the DG Tau jet, and the inferred disk size and extinction found consistent with disc properties from dust continuum maps \citep[see][and Paper I]{Lavalley1997,Pyo2003}. 

 To give a sharper view of the morphology, the raw image was deconvolved by the continuum image, used as an estimate of the contemporary point spread function (PSF).  We used the LUCY restoration routine implemented in the STSDAS/IRAF package. The derived image did not change significantly after 40 iterations. The advantage of deconvolution is mainly to suppress the strong extended non-Gaussian wings of the PSF caused by partial AO correction. The final resolution is not precisely known but, given our limited spatial sampling, is probably close to our Gaussian PSF core before deconvolution (0\farcs12 FWHM).

The deconvolved image of the blue lobe is presented in the bottom left panel of Fig.~\ref{fig:channelsmap}. Bright H$_2$ 1-0 S(1) 
emission appears confined to a small roundish region of about 0\farcs5 = 75 AU in diameter. 
The emission is clearly limb-brightened,  suggesting a "hollow cavity" geometry. 
It is brighter on the side facing the star, and extends sideways up to $\pm 0\farcs2$ from  the jet axis before closing back. A bright peak is also seen along the jet axis at $\Delta Y \simeq -0\farcs1$, and a fainter one at $\Delta Y \simeq -0\farcs4$. Note that we may be underestimating the flux within $<0\farcs1$ of the star due to our procedure for continuum and telluric removal (see Section 2). With this limitation in mind, the morphology of our deconvolved 1-0 S(1) image is strikingly similar to the FUV image of Ly$\alpha$-pumped H$_2$ obtained in July 2011 by \citet{Schneider2013b}, and confirms their conclusion that the H$_2$ wide-angle emission in DG Tau appears essentially stationary over 6 years. We find that the bright H$_2$ rims near the base expanded by less than 0\farcs1 between the two epochs, implying a proper motion expansion speed \Vexp\ $< 12$ \kms. Dedicated monitoring in the same instrumental setup would be necessary to get a more accurate proper motion value. Nevertheless, the current limit already sets tight constraints on scenarios where the H$_2$ emission would originate from the interaction of a wide-angle wind with circumstellar gas (see Section~\ref{sec:discussion}).  

\begin{figure}
\centering 
\includegraphics[width=0.5\textwidth]{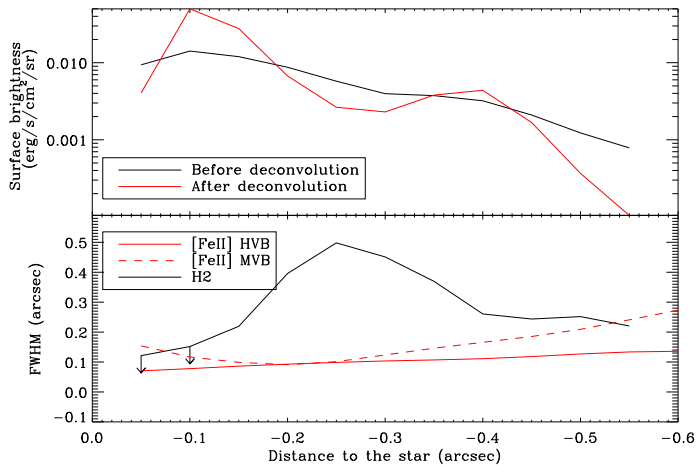}
\caption{{\it Top panel:}  Cut of H$_2$ 1-0 S(1) surface brightness along the blue jet axis, before and after deconvolution by the PSF. The drop in flux inside -0\farcs1 results from our continuum subtraction procedure. Deconvoled peak values should be viewed with caution. 
{\it Bottom panel:} Transverse FWHM of H$_2$ 1-0 S(1) emission in the blue lobe as a function of distance to the star, as measured in the deconvolved image.  In red, we superpose the FWHM of [Fe~II]1.64 $\mu$m at the same epoch in two velocity ranges: MVB (-160 \kms\ $< V <$ +50 \kms) and HVB ($V < -160$ \kms), taken from Paper I.
}
\label{fig:fwhm}
\end{figure}

In the top panel of Fig.~\ref{fig:fwhm} we plot a cut of the surface brightness of H$_{2}$ 1-0 S(1) emission along the blue jet axis as a function of distance from the source, both before and after deconvolution by the continuum.  
The overall flux decrease with distance is very similar to that observed in H$_2$ FUV emission  \citep[see Fig.~4 of][]{Schneider2013b}. The ratio of 1-0 S(1) to FUV H$_2$ lines thus appears roughly constant with distance. The mean brightness level towards the cavity center remains about the same before and after deconvolution, $\simeq 3 \times10^{-3}$ erg s$^{-1}$ cm$^{-2}$ sr$^{-1}$ and is therefore very robust. The peak brightness is more uncertain, since deconvolution has a bias to enhance peaks at the expense of extended lower-level emission beyond -0\farcs4 from the source. We will use the peak brightness before deconvolution as a safe lower-limit.

In the right-hand panel of Fig.~\ref{fig:channelsmap}, we compare the H$_2$ 1-0 S(1)
morphology with {\em contemporaneous} images  from Paper I of the atomic jet in [Fe II]1.64$\mu$m in medium-velocity (MV:  -160 \kms\ $< V <$ 120 \kms) and high-velocity (HV:  $V < -160$ \kms\  and $V > 120$ \kms) intervals. In the red lobe, H$_2$ follows well the base of the bowshock feature seen in the MV range of the [Fe II] emission, suggesting that it traces the low-velocity non-dissociative wings of this bowshock.
In the blue lobe, in contrast, the bright rims of H$_2$ emission are much broader than the [Fe II] jet, that they largely encompass. This is further quantified in  
the bottom panel of Fig.~\ref{fig:fwhm}, where we plot the transverse FWHM of the H$_{2}$ 1-0 S(1) emission in the blue lobe, as measured on the deconvolved image, as a function of distance from the source. The width at $\simeq$ -0\farcs1 is dominated by the bright inner peak, which is only marginally resolved transversally. Beyond this point, the width increases up to 0\farcs5 at a distance of -0\farcs25, indicating an apparent half-opening angle $\simeq$ 45\degree\ similar to that determined by \citet{Schneider2013b} from their FUV image. This is much wider than the 14\degr\ and 4\degr\ measured {\em at the same epoch} for the MV and HV channel maps of [Fe II] (see Fig.~\ref{fig:fwhm} and Paper I). At larger distances, the roundish shape of the H$_2$ emission makes its FWHM decrease again, while the [Fe II] jet keeps its narrow opening.

 In contrast to the marked difference in width between H$_2$ and [Fe II], \citet{Beck2008} noted that the V-shaped H$_2$ emission in DG Tau appears morphologically similar to the low-velocity blueshifted [S II] emission mapped in 1999 with HST by \citet{Bacciotti2000}. \citep{Schneider2013b} noted that the latter actually appears to fill-in the "voids" in the limb-brightened FUV H$_2$ emission, and concluded that the H$_2$ "cavity" is not really hollow but filled by low-velocity atomic flow. A limitation to this argument is that the [S II] HST maps were taken 6-12 years before the H$_2$ near-IR and FUV images, respectively, while the crossing time at 50 \kms\ from the axis to the H$_2$ rim is only 3 years. Data obtained closer in time are thus necessary to confirm this conclusion. In Section~\ref{sec:pvtrans} we present such a comparison both in space and velocity using transverse long-slit [O I] data acquired by \citet{Coffey2007} only 2 years before our SINFONI data.

\subsection{H$_2$ 1-0 S(1) kinematics}

\subsubsection{2D velocity map and 1D velocity gradients along the jet}
\label{sec:pvh2}

\begin{figure}
\includegraphics[trim=0 0 3cm 0,clip,width=0.5\textwidth]{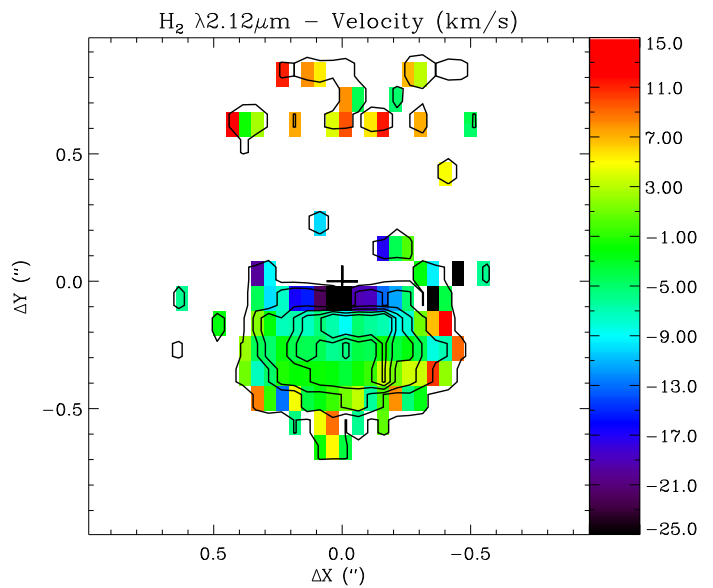}
\caption{Colour map of centroid velocity of the H$_{2}$ 1-0 S(1) line resulting from a one-component Gaussian fit to each individual spectrum, and corrected for uneven slit illumination. Only spaxels with S/N $> 5$ are shown. The cross marks the position of the star. Contours of S/N = 5, 10, 15, 25, 35, 45 are superposed in black ; the fitting error due to noise is $1 \sigma \simeq$ (40/S/N) \kms\ (see text and Section 2).}
\label{fig:centroid_maps}
\end{figure}

Fig.~\ref{fig:centroid_maps} presents a 2D map of the centroid velocities for exposure 2, computed as detailed in Section 2. Contours of S/N are superposed to quantify the centroid Gaussian fitting error in each spaxel. This term generally dominates over the uncertainties due to instrumental drifts from spaxel to spaxel (1.5 \kms\ rms), uneven-slit illumination correction ($\le 2$ \kms), and absolute calibration ($< 2$ \kms). Consistent results are found for exposure 1, considering the slightly different spaxel positions with respect to the central source. 

 Fig.~\ref{fig:centroid_maps} shows that most of the area in the blue lobe is at low radial velocity, with a typical value $\simeq$ -5 \kms\ towards the cavity center. This is in line with the low median velocity barycenter of -2.4 \kms\ reported by \citet{Beck2008} in their comparable spectro-imaging study. At the same time, we find a spatially limited region of higher blueshifts close to the star that appears consistent with previous long-slit data in this region. The highest blueshift of $\simeq -25 \pm $3 \kms\  (at $\Delta Y = -0\farcs07$ and $\Delta X < \pm 0\farcs1$) is in excellent agreement with the $-30 \pm 4$\kms\ observed at the same position in FUV H$_2$ lines in a 0\farcs2 wide slit \citep{Schneider2013a}. SINFONI spectra integrated over the same slit aperture as \citet{Takami2004} also show centroids compatible 
with their reported radial velocity in 1-0 S(1) of $-15 \pm 4$\kms\ (namely: $-8.5 \pm 3$ \kms\  in exposure 1 and $-7\pm 4$ \kms\ in exposure 2, including errors in absolute calibration and uneven slit correction).  The only apparent difference with long-slit data is at $\Delta Y \simeq -0\farcs3$, where our centroid on-axis $\simeq -5 \pm 4$ \kms\ is significantly lower than the $-20 \pm 6$\kms\ reported by \citet{Schneider2013a}. This may be understood by noting that the narrow 0\farcs2 HST slit is dominated by a small axial H$_2$ knot (cf. the FUV image of \citet{Schneider2013b}), while our velocity centroids are measured on the raw (not deconvolved) SINFONI datacubes with large non-gaussian PSF wings, and are thus heavily contaminated by the much brighter H$_2$ rims at wider angles. The axial H$_2$ knot could also be time variable. Overall, our data show that {\em most of the wide-angle H$_2$ emission towards the blue lobe is at significantly lower radial velocity than inferred from long-slit spectra}.

 In order to increase precision on velocity gradients along the flow and in the redshifted lobe, we increased S/N by building for each exposure a Position-Velocity (PV) diagram along the jet in a pseudo-slit of 1\arcsec~width, keeping the original $Y$-sampling along the jet axis defined by the positions of the SINFONI slicing mirrors. The faint spectra at distances farther than -0\farcs7 (blue side) and between the star and +0\farcs5 ("dark lane" on the red side) were further averaged to increase signal to noise. We carried out the analysis individually for each exposure; since their corrections for absolute calibration and uneven slit-illumination differ (see Section 2), comparison of the results gives a useful cross-check, while also improving spatial sampling of velocity gradients along the jet axis, where the slicing mirror width (0\farcs1) undersamples the core of the PSF (0\farcs12 FWHM).

\begin{figure*}
\sidecaption
\includegraphics[width=0.7\textwidth]{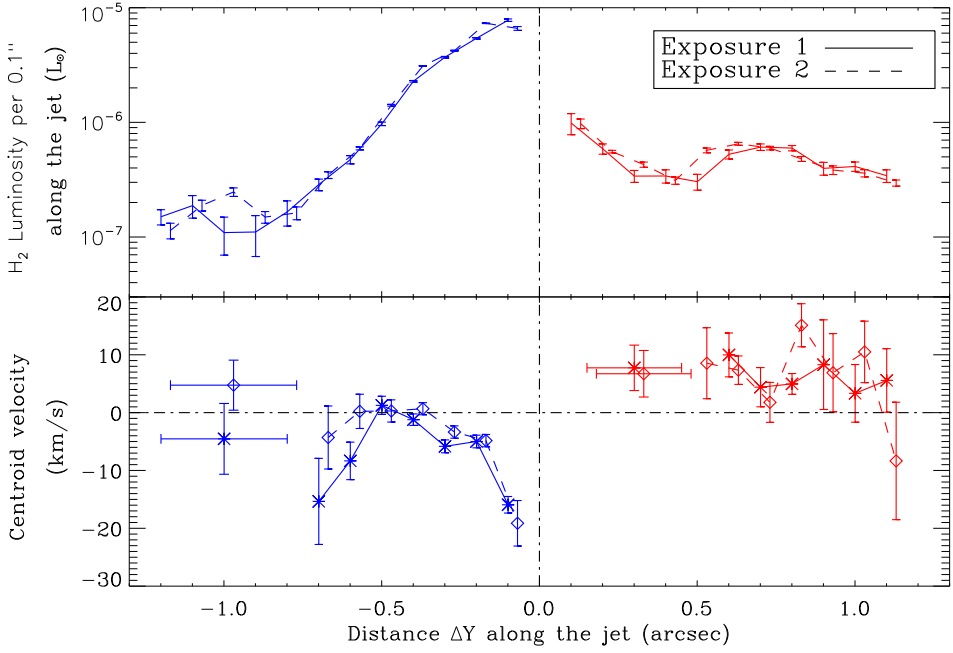}
\caption{{\it Top:} Luminosity of H$_2$ 1-0 S(1) per 0\farcs1 length along the jet,  integrated over 1\arcsec\  across the jet, for each exposure. {\it Bottom:}  Gaussian-fitted centroid velocity of H$_2$ 1-0 S(1) in spectra averaged across a 1\arcsec\ wide pseudo-slit. Correction for uneven slit-illumination was applied to each averaged spectrum (see Appendix~\ref{app:uesi}). Absolute velocity calibration is better than 2 \kms\ in each exposure. Vertical $1\sigma$ error bars include the fitting error due to noise, and 0.8\kms\ rms instrumental calibration drifts between slicing mirrors  (see Section 2).  Horizontal error bars indicate where spectra were further averaged along the jet to increase S/N. }
\label{fig:pvh2}
\end{figure*}

The top panel of Fig.~\ref{fig:pvh2} plots the luminosity inside each "slice" of the PV diagram. 
The bottom panel of Fig.~\ref{fig:pvh2} plots the H$_2$ centroid velocities fitted to spectra averaged across the jet as a function of distance along the jet, for each exposure separately.  The two individual exposures show excellent agreement within the error bars, giving us confidence in our velocity determinations and error estimates.  The velocity gradient along the blue lobe seen in Fig.~\ref{fig:centroid_maps} is confirmed: The 1D-averaged $H_2$ centroids show an apparent deceleration from $\simeq -18$ \kms\  at $\Delta Y \simeq -0\farcs1$ to $\simeq -5$ \kms\ at $-0\farcs2$, and to essentially zero near the cavity end at $\Delta Y = -0\farcs5$. This gradient is clearly too strong to be caused by measurement uncertainties; residual errors in uneven slit illumination correction are at most 2\kms\ within -0\farcs3 from the star, and appear negligible further out (cf. Appendix~\ref{app:uesi}), while residual errors in absolute calibration would only shift each curve without affecting the gradients. 

In the fainter region beyond the cavity edge, both exposures show a trend for an increased blueshift at $\simeq -10$ \kms\ at $\Delta Y \simeq$ -0\farcs6, that might be tentative evidence for a more distant faint H$_2$ knot along the jet axis. This position corresponds to the tip of the bright [Fe~II] HVC jet beam in Fig.~\ref{fig:channelsmap}. Even fainter H$_2$ emission is detected around -1\arcsec, with centroid velocities close to zero. A faint bubble feature is seen at this position in [Fe~II] HV and MV (see Fig.~\ref{fig:channelsmap} and Paper I). Finally, the H$_2$ 
emission in the red lobe appears mostly redshifted, with a mean radial velocity $\simeq$ +5 \kms. In the rest of this paper, we focus on the bright wide-angle H$_2$ cavity in the blue lobe, as the remaining H$_2$ features have too low S/N for more detailed analysis.


\subsubsection{Comparison with kinematics in atomic lines}
\label{sec:pvtrans}

The velocity range of [Fe II]1.64$\mu$m emission in the blue lobe of DG Tau appears totally distinct from that of H$_2$. The [Fe II]1.64$\mu$m SINFONI spectra in the region of the H$_2$ cavity have centroids ranging from -150 to -200 \kms\ (Paper I). A higher resolution long-slit [Fe II]1.64$\mu$m spectrum obtained in 2001 by \citet{Pyo2003} reveals a separate lower-velocity component peaked around -50 to -100 \kms, but still very weak emission in the velocity range of H$_2$ 1-0 S(1), and an apparent acceleration away from the star in opposite sense to the centroid gradient observed in H$_2$. Together with the much narrower opening angle of [Fe~II] compared to H$_{2}$ 1-0 S(1) 
(cf. Sect.~\ref{sec:map}) this appears to suggest no direct dynamical interaction between the [Fe~II]  jet and the wide-angle H$_2$ in DG Tau. The same kinematic and spatial separation between [Fe II] jet and H$_2$ wide angle emission was observed previously in the case of HL Tau by \citet{Takami2007}, who then interpreted the V-shaped H$_2$ as a shocked outflow cavity swept-up by an unseen wide-angle wind.

\begin{figure}
\centering
\includegraphics[width=0.5\textwidth]{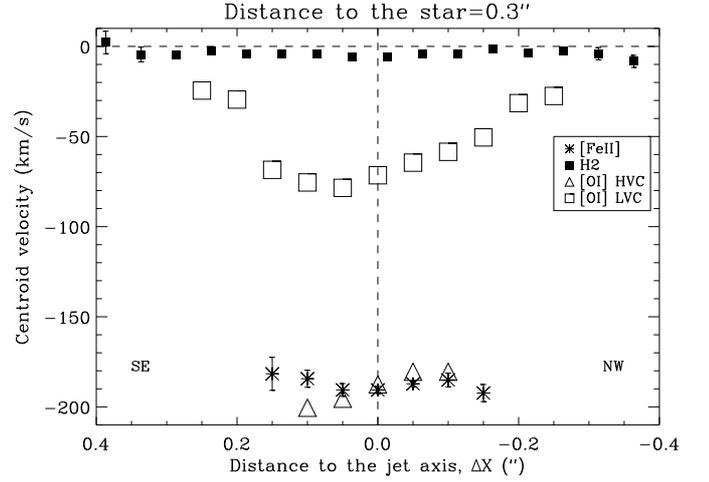}
\caption{Centroid velocity versus distance from the jet axis for H$_{2}$ 1-0 S(1) 
(filled squares, this paper), [FeII] 1.64 $\mu$m (stars, from Paper I) and [OI]$\lambda$6300{\AA} HVC and LVC \citep[ open symbols, from][]{Coffey2007}), all measured along a 0\farcs1-wide transverse cut at $\Delta Y = -0\farcs3$ from the star into the blue lobe. Positive $\Delta X$ is to the SE. }
\label{fig:pv_trans_feii-h2-oi}
\end{figure}

In the case of DG Tau, previous optical spectro-imaging in [O~I] and [S~II] do provide direct evidence for  a wide-angle atomic wind, with a decreasing velocity from axis to edge \citep{Lavalley-Fouquet2000,Bacciotti2000,Bacciotti2002}.
However, these optical data were taken 6-7 years before the SINFONI H$_2$ data, at a time when the DG Tau jet exhibited very high-velocity emission up to -350 \kms\ that was not observed since then. 
This makes a comparison between the two datasets somewhat uncertain. 
In order to investigate the spatio-kinematic relationship between the wide-angle H$_2$ and the atomic flow using data as close in time as possible,  we compare in Fig.~\ref{fig:pv_trans_feii-h2-oi} the transverse velocity structure at $\Delta Y = -0\farcs3$ across the blue jet in the 1-0 S(1) line (from this paper), in the [Fe~II] 1.64 $\mu$m line observed simultaneously (from Paper I) and in the optical [O I]$\lambda$6300\AA\ line observed only 2 years earlier with HST/STIS \citep[from][]{Coffey2007}. The [O I] profiles were fitted by two gaussians by \citet{Coffey2007}: a "high-velocity component" (HVC) and a "low-velocity component" (LVC). Both centroids are reported in the figure. The [Fe II] profiles are narrower and under-resolved by SINFONI. They were thus fitted by single gaussians, dominated here by the HVC.

Figure \ref{fig:pv_trans_feii-h2-oi} shows that while the [O I] HVC coincides in velocity and spatial width with the narrow [FeII] HVC, the [O I] LVC is much slower and spatially broader; its radial velocity drops from -60 \kms on-axis to -25 \kms~at $\pm$0\farcs25 from the axis, and the spatial FWHM is $\simeq 0\farcs2$ \citep{Coffey2008}. Therefore, the [O~I] LVC is filling-in the "gap" between [Fe II] and H$_{2}$ emission, both spatially {\em and} kinematically. In addition, we note that the centroid variations and spatial FWHM of the [OI] LVC observed in 2003 by \citet{Coffey2008} are identical to measurements obtained 4 years earlier at the same distance from the star \citep[see Fig.~1 in][and Fig.~14 in Maurri et al. 2013]{Bacciotti2002}. We thus find that the spatio-kinematic structure of the LVC remains remarkably stable over more than a transverse crossing time, despite strong variability of the HVC at velocities $\ge 300$\kms. This finding definitely strengthens conclusions based on earlier 1999 HST data that the wide-angle H$_2$ cavity in DG Tau appears as a slower, outer extension of the ``onion-like" velocity structure of the atomic flow \citep{Takami2004,Beck2008, Schneider2013b}.

Another important implication of this comparison is that {\em [Fe II] emission does not appear as a good tracer of the wide angle atomic flow slower than 50 km/s}. While Pyo et al. detect [Fe II] LVC emission within $\pm 0.15''$ of the jet axis at velocities bluer than -50 km/s, the comparison of transverse PV diagrams in [Fe II] and [O I] (See Fig.~5 of Paper I) shows no detectable contribution in [Fe II] from the wider angle, slower gas below -50 \kms~that so clearly stands out in [O I] (and [S II]). This difference cannot be explained by gradients in density or temperature: the [Fe II]1.64$\mu$m line has a critical density intermediate between that of [O I] and [S II], and a similar upper level energy. We propose that it is due instead to iron being increasingly depleted at lower speeds. Indeed, modeling of the [Fe II] to [O I] line ratio in Paper I indicates a higher average gas-phase iron depletion of a factor 10 in the interval -100 to +10 \kms, compared to only a factor 3 in the interval -250 to -100 \kms. Depletion could be so high in the velocity range -50 to -25 \kms\ that the corresponding wide-angle emission seen in [O I] and [S II] remains undetectable in [Fe II] at the sensitivity and resolution of SINFONI.

\subsubsection{Constraints on H$_2$ rotation}
\label{sec:vshift}

\begin{figure}
\centering
\includegraphics[width=0.5\textwidth]{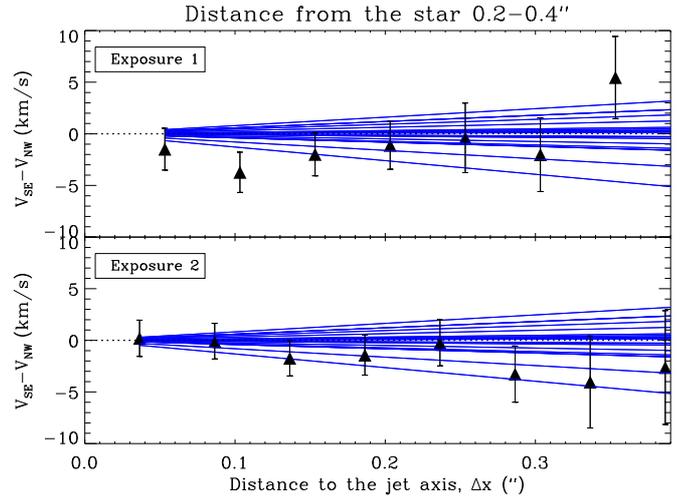}
\caption{ Velocity shifts in H$_2$ 1-0 S(1) between symmetric positions on either side of the jet axis, after averaging across 3 slicing mirrors with $-0\farcs4 \le \Delta Y \le -0\farcs2$. 
The $1\sigma$ error bars include the fitting error due to noise and the 1\kms\ rms instrumental drifts from spaxel to spaxel. Blue lines illustrate the typical range of spurious instrumental velocity gradients along slicing mirrors, as measured from OH lines (see Section 2). Residual errors in absolute calibration and correction for uneven-slit illumination cancel out here. 
}
\label{fig:vshift}
\end{figure}


 The LVC centroids in optical forbidden lines are known to display clear asymmetries between opposite sides of the jet axis, with amplitude and sign consistent with a rotating, steady magneto-centrifugal disk wind launched out to $\simeq$ 3 AU \citep[][cf. Fig.~\ref{fig:pv_trans_feii-h2-oi}]{Bacciotti2002, Anderson2003, Pesenti2004, Coffey2007}. We therefore make use of our SINFONI data to search for transverse velocity gradients possibly indicative of rotation in the wide-angle H$_2$ lobe.

The differences in centroid velocity of H$_2$ 1-0 S(1) between symmetric positions across the jet axis are represented in Fig.~\ref{fig:vshift}. Note that any residual errors in absolute calibration and uneven-slit illumination correction cancel out  (the illumination is symmetric with respect  to the jet axis, to first order). To increase S/N, we averaged the 3 slicing mirrors centered at distances $\Delta Y$ between $\simeq$0\farcs2 and $\simeq$0\farcs4 from the star, where H$_2$ emission is broadest. We then determined the velocity centroid by Gaussian fitting at each offset $\Delta X$ across the axis (keeping the original spaxel sampling of 0\farcs05 in this direction). In exposure 2, spaxels are not positioned symmetrically on either side of the jet axis, therefore centroid velocities at $\Delta X < 0$ were lineary interpolated at positions symmetric from those at $\Delta X > 0$ before computing velocity differences. Individual results from the two exposures are shown in Fig.~\ref{fig:vshift}. 
Blue lines superposed in Fig.~\ref{fig:vshift} illustrate the expected range of spurious velocity shifts due to instrumental effects along slicing mirrors ("slope effects", see Section 2). 

When taken individually, all velocity differences plotted in Fig.~\ref{fig:vshift} are compatible with zero to within their respective error bars, except in exposure 1 at $\pm 0\farcs1$ from the axis. However, there is a systematic consistent trend at most positions and in both exposures for spectra at positive $\Delta X$ (SE) to be more blueshifted than those at negative $\Delta X$ (NW). This would correspond to rotation in the same sense as the CO disc and optical jet in DG Tau \citep{Testi2002, Bacciotti2002, Coffey2007}. Unfortunately, the observed trends are also of the same order as the maximum residual instrumental slope along slicing mirrors observed in OH sky lines (blue lines in Fig.~\ref{fig:vshift}). Although (i) most instrumental residual slopes are  shallower than this, (ii) averaging over 3 slicing mirrors should decrease the effect, and (iii) there is only a 25\% probability that it would go in the same sense as the observed gradient in both exposures, an instrumental origin cannot be totally excluded. Since we lack a bright OH sky exposure simultaneous with our data to correct for the effect spaxel-by-spaxel, we can only put an upper limit on the true transverse velocity shifts in H$_2$ present in our data. 

 To obtain a conservative upper limit on the possible rotational shift at $\Delta X \pm 0\farcs2$ from the axis (where the H$_2$ rims peak, cf. Fig. 1), we take the average shift of the two exposures at this offset (-1.3\kms), add its $3\sigma$ error bar (-4 \kms), and correct by the maximum expected instrumental effect in the  {\em opposite} direction $\simeq$ 2 \kms (since this situation cannot be excluded either). We obtain an upper limit of $V_{\rm shift} \le 7$ \kms\ in the sense of disc rotation. In the ideal situation of a single flow streamline observed with infinite angular resolution, this would set an upper limit on the H$_2$ flow rotation speed  of $V_{\phi} = V_{\rm shift}/(2\sin i) \le 5$\kms\ for a jet inclination to the line of sight of 40$^{\circ}$. In reality, convolution by the PSF and contribution from several streamlines along the line of sight both act to reduce the observable velocity shifts \citep{Pesenti2004}, and the true $V_{\phi}$ could be somewhat higher than this. 
 

\subsection{H$_2$ temperature, column and volume densities, and emitting area}
\label{sec:h2column}

\cite{Beck2008} computed the ratios of various ro-vibrational lines of H$_2$ towards the H$_2$ peak position, and found that they were compatible with a thermalized population at $T \simeq 2000$~K. To search for possible temperature gradients, we computed the 2-1 S(1)/1-0 S(0) ratio at several positions, after averaging over 0\farcs2 along the jet and 0\farcs5 across the jet to increase signal to noise in the fainter $v$ = 2--1 line. We found a similar ratio at $\Delta Y=$ -0\farcs2 and -0\farcs4  along the blue jet, of 0.07 $\pm$ 0.02 and 0.09 $\pm$ 0.02 respectively, consistent with the $0.07\pm 0.03$ reported by \cite{Beck2008} towards the H$_2$ peak. Hence we assume a uniform temperature $T \simeq$ 2000~K in the following. We note that the 1-0 S(0)/1-0 S(1) ratio  $\simeq 0.25 \pm 0.03$ observed towards the H$_2$ peak \citep{Takami2004, Beck2008, Greene2010} combined with Eq.~5 of \citet{Kristensen2007} indicates an H$_2$ ortho/para ratio $\simeq$ 3 consistent with the LTE value. We assume the same holds throughout the region. 


Our flux-calibrated data in H$_{2}$ 1-0 S(1) 
 can be converted into column densities in the upper level of the transition ($v$=1, $J$=3) through $N_{1,3}= F_{1-0S(1)}\times (4\pi)/(h\nu_{ul}A_{ul})=3.8\times10^{19}$ \cmsq\ $\times F_{1-0S(1)}$(\mbox{erg s$^{-1}$ cm$^{-2}$ sr$^{-1}$}). Assuming LTE at 2000 K and an ortho/para ratio of 3, the total column density of hot H$_{2}$ gas is then given by $N_{H_{2}}=N_{1,3}/1.28\times10^{-2}$ \citep{Takami2004}. The inferred column density of hot H$_2$ reaches $4\times 10^{19}$ \cmsq\ towards the H$_{2}$ peak at -0\farcs1 from the star (peak brightness $\simeq 0.013$ \mbox{erg s$^{-1}$ cm$^{-2}$ sr$^{-1}$}; cf. Fig.~\ref{fig:fwhm}), while the typical value at the center of the blue lobe is  $\simeq 10^{19}$ \cmsq\ (central brightness $\langle F_{1-0S(1)} \rangle \simeq 3 \times10^{-3}$ erg s$^{-1}$cm$^{-2}$ sr$^{-1}$; cf. Fig.~\ref{fig:fwhm}).  Since the inner peak might have a contribution from an axial jet knot \citep[see FUV image of][]{Schneider2013b}, and the rims are clearly limb-brightened, we consider the brightness towards the blue lobe center as our best estimate for the intrinsic "face-on" column density of the layer producing the wide-angle H$_2$ emission. Both our image and that of \citet{Schneider2013b} suggest that the rims are thinner than 0\farcs1 = 14 AU, yielding a lower limit to the volume density of hot H$_2$ in the emitting layer of $n_{\rm H_2} \ge 5 \times 10^4$ \cm.

The top panels of Fig.~\ref{fig:pvh2} plot the 1-0 S(1) luminosity {\it per 0\farcs1 length} along the jet, integrated over a full width of 1\arcsec across the jet, for each individual exposure. Even though the H$_2$ emission is narrowest at the base and widens out, the luminosity still peaks at -0\farcs1 and then decreases gradually with distance.  The total luminosity integrated over the whole emitting region is L$_{1-0S(1)}$ = 1.9$\times$10$^{-5}$ L$_{\sun}$, only 50\% more than in the 0\farcs3-wide slit of \citet{Takami2004}. The total mass of hot H$_{2}$ at 2000 K, assuming LTE, is then 3$\times$10$^{-8}$ M$_{\sun}$, while the characteristic emitting area at average brightness level is $A \simeq L_{1-0S(1)}/(4\pi \langle F_{1-0S(1)} \rangle)$ = $2 \times 10^{30}$ cm$^{2} \simeq$  (90 AU)$^2$.

We note that we do not include any correction for dust attenuation of the H$_2$ 1-0 S(1) line fluxes. Although \citet{Beck2008} derived a rather high $A_V = 11.2\pm 10$ mag towards the H$_2$ peak, the uncertainty in their estimate is large, due to strong telluric absorption in H$_2$ Q-branch lines near 2.4 $\mu$m. Furthermore, the optical jet brightness keeps increasing all the way up to 0\farcs1 from the star \citep{Bacciotti2000}, suggesting that the atomic jet suffers only moderate circumstellar dust extinction. Hence, any 11 mag obscuring layer would have to lie behind the blue atomic jet lobe yet in front of the blueshifted H$_2$ peak, which seems very contrived. 
Finally, the strikingly similar morphology of FUV H$_2$ emission imaged by \citet{Schneider2013b} to that seen in the near-infrared 1-0 S(1) line also argues strongly for a low A$_V$ to the hot H$_2$.  We thus assume that dust obscuration is negligible in K-band lines.

\section{Origin of the wide-angle H$_2$ emission in DG Tau}\label{sec:discussion}

In this section, we re-examine possible origins for the wide-angle, limb-brightened H$_2$ 1-0 S(1) emission in the blue lobe of DG Tau, taking into account the new information extracted from our SINFONI data on the geometry, proper motion, radial velocities, and surface brightnesses (cf. previous Section). In subsection~\ref{sec:shocks}, we determine the allowed shock parameter space able to explain both the 1-0 S(1) surface brightness and the 2-1/1-0 S(1) line ratio. In subsection~\ref{sec:toymodels}, we model geometrical and projection effects to constrain the velocity field compatible with the observed radial velocities. Finally, we combine these two sets of constraints to examine the plausibility and requirements of several scenarios proposed for the origin of wide-angle H$_2$ in DG Tau and similar sources (e.g., HL Tau):  
\begin{itemize}
\item an irradiated disk atmosphere / envelope 
\item a molecular MHD disk wind heated by ambipolar diffusion 
\item a {\em forward} shock driven into the disc/envelope 
\item a {\em reverse} shock driven back into a molecular wide-angle wind 
\end{itemize}



\subsection{Allowed shock parameter space}
\label{sec:shocks}
\citet{Beck2008} compared the H$_2$ line {\em ratios} in DG Tau with planar shock models from \citet{Smith1995} with preshock density $2 \times 10^6$\cm. They mention good agreement with both an 8 \kms\ non-dissociative J(ump)-type shock, and a 35 \kms\ multifluid C(ontinuous)-type shock (where ions and neutrals are decoupled).  We here use  the additional constraint provided by the calibrated 1-0 S(1) H$_2$ surface brightness, together with a larger grid of planar shocks computed by \citet{Kristensen2008}, to explore more extensively the shock parameter space that could reproduce the DG Tau observations. 

The predicted H$_{2}$ surface brightnesses depend on three parameters: the shock velocity \Vs, the preshock hydrogen nucleus density, \nH, and the dimensionless magnetic field parameter\footnote{This parameter is related to the Alfv\' en speed in the preshock gas through $V_{\rm A} = b \times1.88$\kms.} $b\equiv B(\mu{\rm G})/\sqrt{n_{\rm H}({\rm cm}^{-3})}$, where $B$ is the magnetic field component parallel to the shock front. The grid includes both single-fluid J-type shocks with $b = 0.1$, and steady multifluid C-type shocks with $b$ ranging from 0.5 to 10. Preshock densities range from $10^4$ to $10^7$ \cm. The preshock gas is assumed fully molecular with no UV field and a standard ISM dust content and cosmic ray flux; the latter determines in particular the initial abundances of charged species and  atomic H. 

We note that model predictions assume a face-on view. A sideways view would increase the observed surface brightness by a factor $1/\cos(i)$ for an infinite slab. This effect does not introduce large errors when fitting H$_2$ emission near the head of a bowshock, where the path-length is limited by the strong curvature \citep{Gustafsson2010}, but will be more severe for a tilted cavity where the front side is almost entirely tangent to the line of sight, as suggested here by the brighter rim towards the star (see Sect.~\ref{sec:toymodels}). Hence, we will rather compare the face-one shock predictions with the 1-0 S(1) surface brightness towards the center of the H$_2$ cavity, where limb-brightening effects will be minimal.  

\begin{figure}
\includegraphics[width=0.5\textwidth]{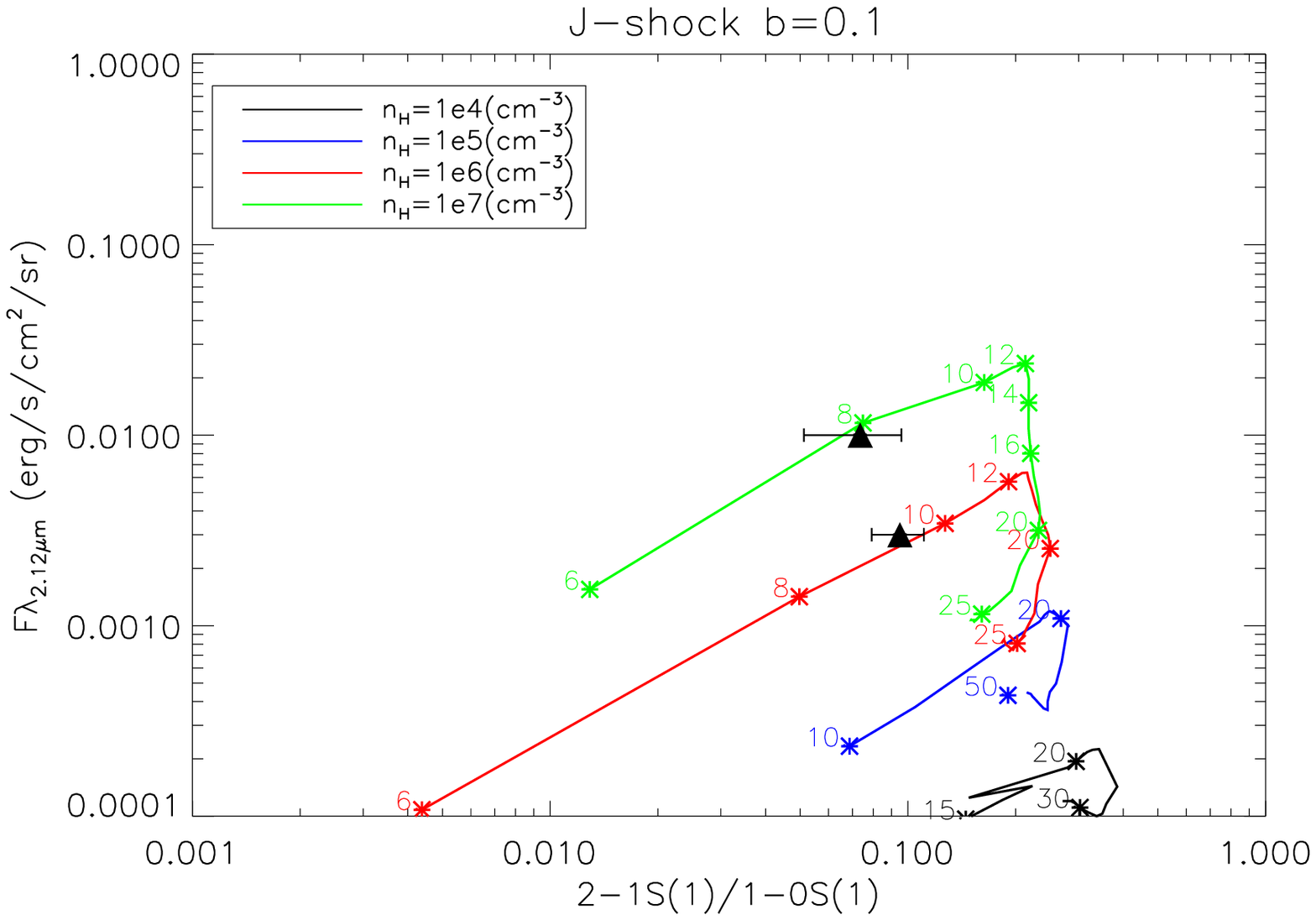}
\includegraphics[width=0.5\textwidth]{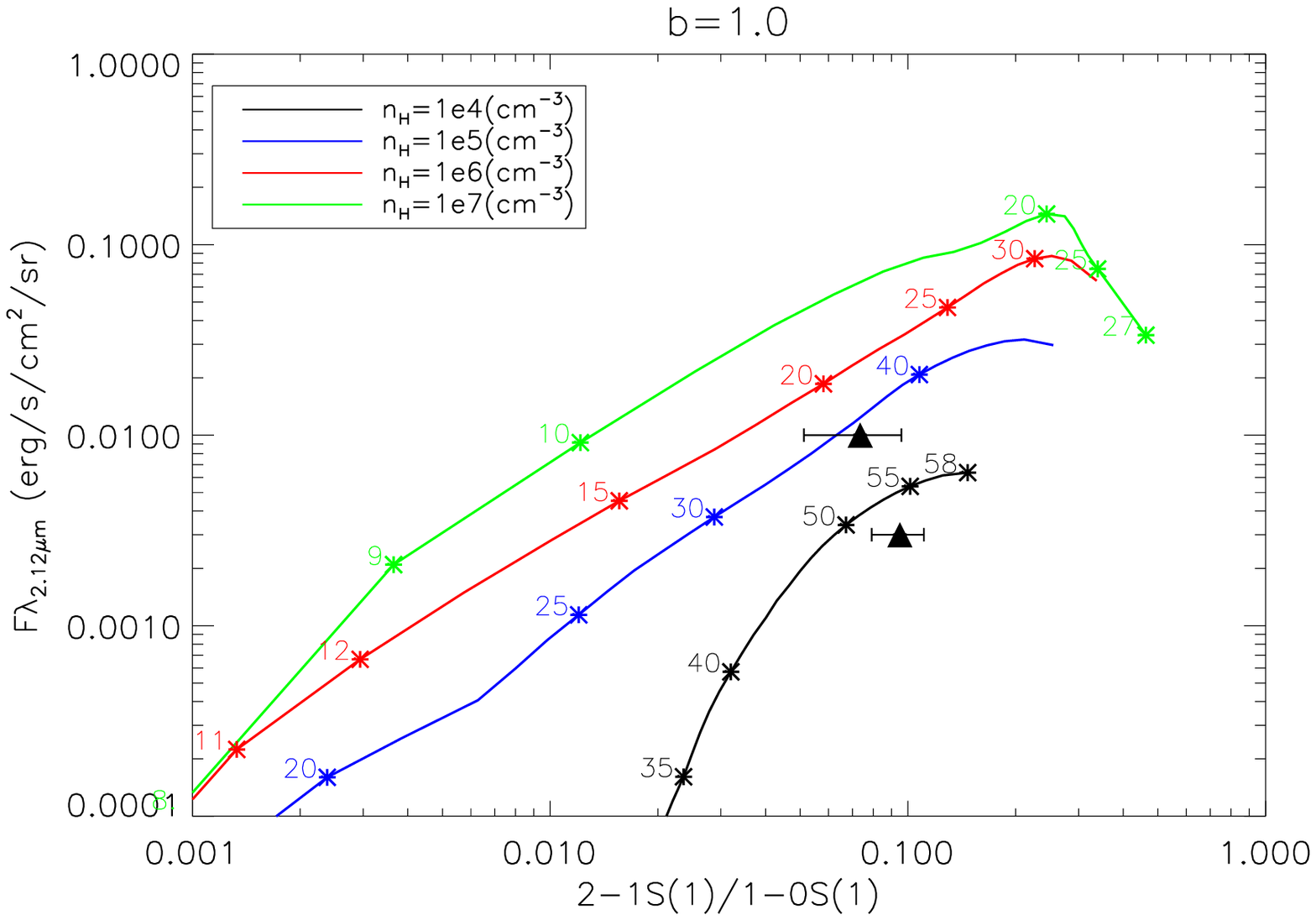}
\caption{H$_{2}$ surface brightness in the 1-0 S(1) line against the $v=$2-1 S(1)/$v=$1-0 S(1) ratio as observed in the DG Tau blue lobe (lower triangle: center of blue lobe, upper triangle: inner peak) and as predicted for planar shock models viewed face-on from \citet{Kristensen2008} (colour curves). The error bars on observed surface brightness are smaller than the symbol size, while the difference between the two symbols illustrates the maximum uncertainty due to projection / limb-brightening effects. Top panel: planar J-shocks with $b=0.1$. Bottom panel: C-shocks with $b=1$. Each model curve corresponds to a different value of the preshock hydrogen nucleus density \nH,  increasing from bottom to top. The shock speed \Vs\ increases to the right and some values are marked along the curves to guide the eye.  Similar plots for C-shocks with $b=0.5$ and $b=10$ are shown in Appendix~\ref{an:shocks_models}. Note the narrow range of shock speeds reproducing the observed 2-1 S(1)/1-0 S(1) ratios. }
\label{fig:molec_model}
\end{figure}

In Figure~\ref{fig:molec_model} we compare the observed 1-0 S(1) surface brightness  and 2-1 S(1)/1-0 S(1) ratio with model predictions for J-shocks with $b=0.1$ (top panel) and C-shocks with $b=1$ (bottom panel). It may be seen that the 2-1 S(1)/1-0 S(1) ratio is much more sensitive to \Vs\ than to \nH, with changes of a factor 2 in the former having a stronger effect on the ratio than a change of a factor 10 in the latter. In the case of J-shocks, the data point towards the cavity center is fitted with a high  \nH $\simeq 10^6$\cm\ and a low shock speed \Vs $\simeq 9-10$\kms. In the case of C-shocks with $b$=1,  the models point to a lower density \nH $\simeq 10^4$\cm\ and a higher shock speed \Vs $\simeq$ 52\kms. Similar plots for C-shocks with $b$=0.5 and $b$=10 are shown in Appendix~\ref{an:shocks_models}. The data point for the inner peak at -0\farcs1 from the star (top triangular symbol in Figure~\ref{fig:molec_model}) would require a roughly 10 times higher preshock density than the central point.  However the required shock speed would change by less than a factor 2. Therefore, our conclusions on \Vs\ are robust even if the surface brightness at the cavity center is affected by residual geometric projection effects.

Table~\ref{tab:shocks} summarizes the pairs of values (\nH,\Vs) compatible with observations for each magnetic field parameter $b$. It may be seen that a larger magnetic parameter $b$  requires higher shock speed and lower density  to match the observations. It is noteworthy that in all cases, the allowed range of shock speed reproducing the observed 2-1 S(1)/1-0 S(1) ratio is very narrow, setting strong constraints on scenarios of shock-heating.
For each planar shock model able to reproduce the H$_2$ 1-0 S(1) brightness and 2-1/1-0 ratio, we also calculated $V_n({\rm H}_2)$, the (emission-weighted) centroid velocity of the H$_2$ 1-0 S(1) emission layer {\em in the reference frame of the shock wave}. In J-shocks, where deceleration is almost instantaneous, $V_n({\rm H}_2)$ is essentially zero. In contrast, we find that it is typically $\simeq$60\% of the shock speed in the C-shocks of Table~1, where deceleration of the neutrals occurs through collisions with the (rare) ions and is therefore much more gradual.

\begin{table*}
\caption{ Planar shock parameters for the peak and central positions in the H$_2$ cavity, assuming a face-on view}
\begin{tabular}{c|c|c|ccccccc}
\hline
\hline
\multicolumn{2}{c|}{  }  & J-shock & \multicolumn{7}{|c}{C-shocks}\\
\cline{3-10}
\multicolumn{2}{c|}{$b = B(\mu{\rm G})/\sqrt{n_{\rm H}({\rm cm}^{-3})}$}  & 0.1 & 0.5  & 1.0    & 2.0   & 3.0  & 4.0   & 5.0   & 10.0\\
\hline
\hline
\multirow{2}{*}{Peak} &  n$_{\rm H}$ (cm$^{-3}$) & 10$^{7}$ & $10^{5}-10^6$   & 5$\times$10$^{4} - 10^5$ & $\sim$4$\times$10$^{4}$ & $\sim$3$\times$10$^{4}$ & $\sim$3$\times$10$^{4}$ & $\sim$2$\times$10$^{4}$ &10$^{4}$ \\
   & V$_s$ (\kms)  &    8     &  32--16   &  41--35        &  $\sim$50    & $\sim$59       & $\sim$65      & $\sim$70    & 88 \\
\hline
\multirow{2}{*}{Center} & n$_{\rm H}$ (cm$^{-3}$) 
& $10^{6}$ & $\simeq 5 \times10 ^{4}$ & $\simeq 10^{4}$ & $\le 10^{4}$  &$\le 10^{4}$  &$\le 10 ^{4}$  &$\le 10^{4}$ & $<10^{4}$ \\
& V$_s$ (\kms)
& 9--10  & $\simeq 30$   & 52       & $\ge$60       & $\ge$65      & $\ge$70       & $\ge$75     & $>$90  \\
\hline
\end{tabular}
\label{tab:shocks}
\end{table*}

\subsection{Constraints on the intrinsic velocity field of the wide-angle H$_2$}
\label{sec:toymodels}

The limb-brightened morphology of the wide-angle H$_2$ emission is suggestive of a tilted cavity. Hence,  observed radial velocities are affected by complex projection effects.  We thus need 3D geometrical models to infer the true velocity field of the H$_2$ emitting gas parallel ($V_\parallel$) and perpendicular ($V_\perp$) to the cavity surface. To remain as general as possible, we consider two basic toy models : (1) a hollow cone with its apex at the source position and its axis aligned with the blue jet, and (2) a (half or full) hollow sphere centered at a projected distance $z_{\rm proj}$ from the source along the blue jet axis. They are assumed infinitely thin and with uniform surface brightness. We built synthetic 3D datacubes for several inclinations to the line of sight. Each channel map is convolved by a gaussian PSF of FWHM 0\farcs12,  and the PSF of the spectrum at (0,0) is subtracted out (as performed in our observations for the purpose of telluric and continuum subtraction, see Section 2). Residual emission maps are then built and  compared by eye with the deconvolved image of Fig.~\ref{fig:channelsmap}, to determine the geometrical parameters reproducing the H$_2$ cavity shape, size, and limb-brightened appearance.

\begin{figure}
\includegraphics[width=0.5\textwidth]{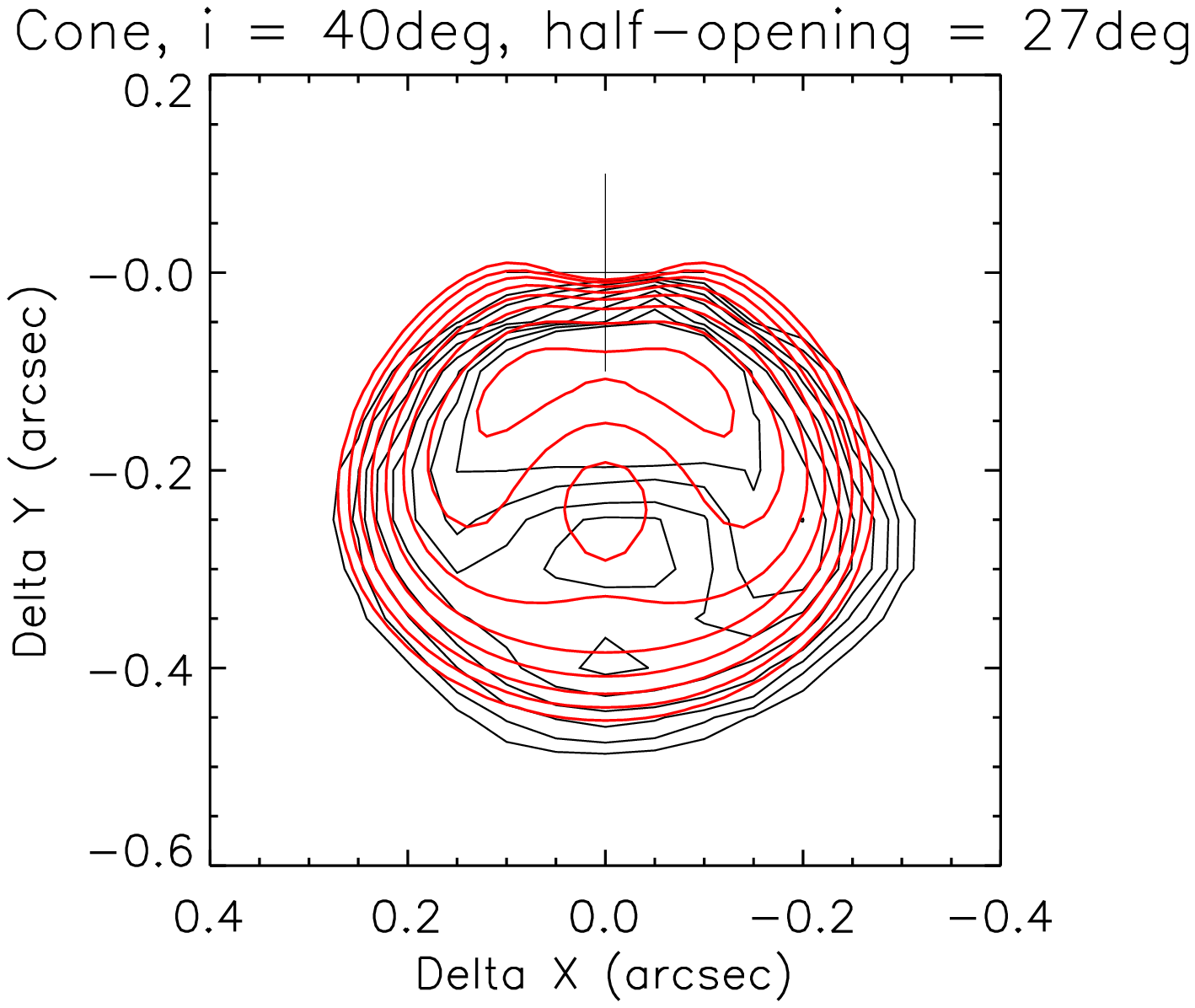}\\
\includegraphics[width=0.5\textwidth]{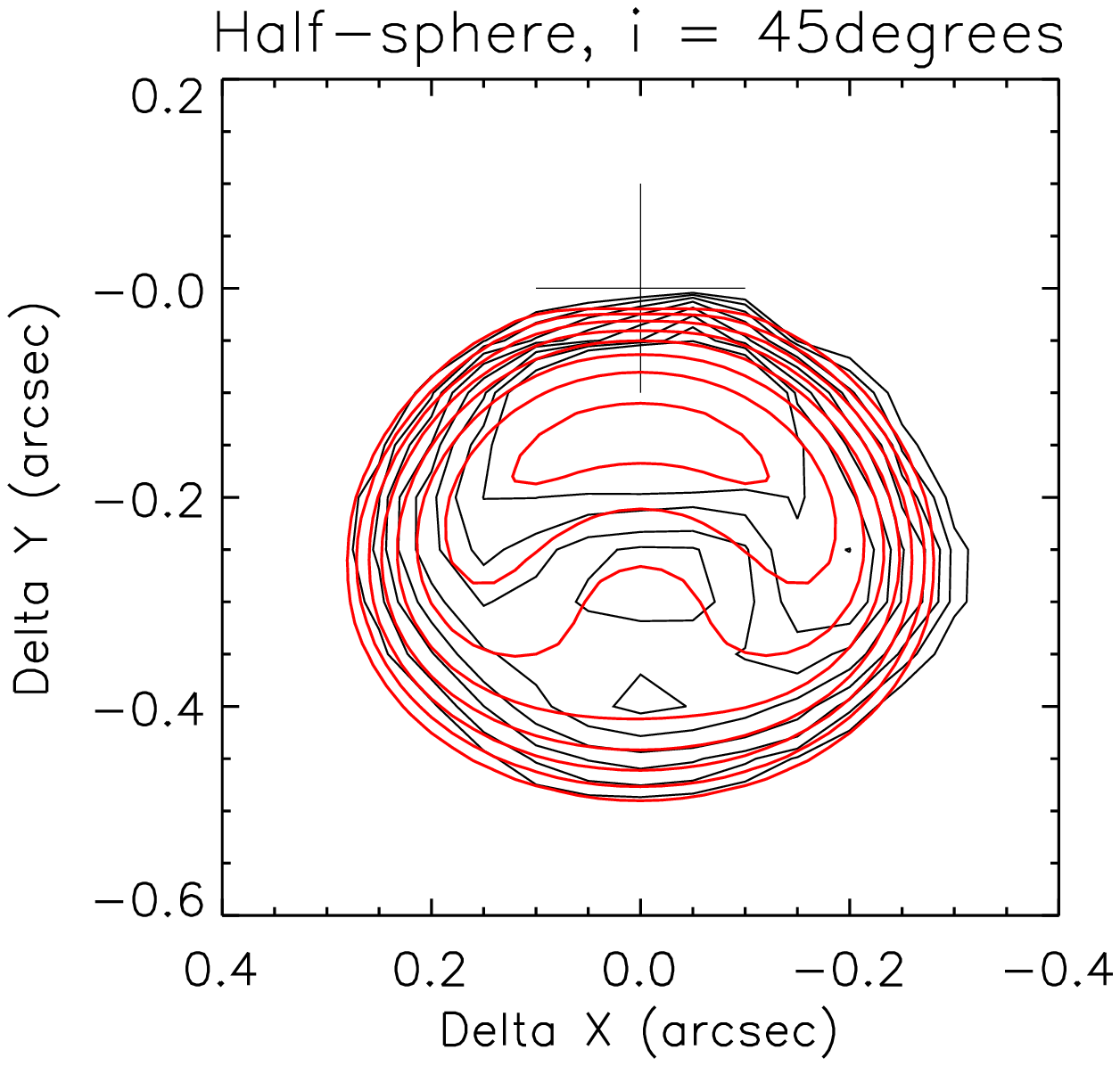}
\caption{ Predicted emission maps for toy models (black contours) convolved by a gaussian of 0\farcs12 FWHM, superposed on the observed deconvolved H$_2$ 1-0 S(1) map from Fig.~\ref{fig:channelsmap} (red contours). {\it Top:} hollow cone viewed 40\degr\ from pole-on with radius 60 AU and semi-opening angle 27\degr. {\it Bottom:}  hollow half-sphere viewed 45\degr\ from pole-on, with radius 28 AU and center projected at 40 AU along the blue jet. All contours increase by factors of $\sqrt{2}$. The PSF of the spectrum at (0,0) was subtracted in all cases. }
\label{fig:cone-map}
\end{figure}

In the case of cones, the  deconvolved map is best reproduced with a spherical radius $R_{\rm max} \simeq$  0\farcs43 (60 AU) and half-opening angle $\theta$ = 30\degr$\pm$3\degr\ for inclinations $i$ = 45\degr$\pm$5\degr  to the line of sight. A typical predicted map for $\theta$ = 27\degr\ and $i$ = 40\degr\ is shown in the top panel of Fig.~\ref{fig:cone-map}, superposed on the observed deconvolved map. The front side of the cavity is almost tangent to the line of sight, so that the limb-brightened rim gets brighter closer to the star, as observed; the narrow peak observed at -0\farcs1 is not fully reproduced, and may include contribution from a jet knot. The cone surface is (75 AU)$^2$, in good agreement with the full emission surface $\simeq$ (90 AU)$^2$ estimated in Section~\ref{sec:h2column}. 

In the case of a sphere, the best fit to the overall size and shape is for a radius $R_{\rm max} \simeq$ 0\farcs2 (28~AU) centered at $z_{\rm proj} \simeq$ 0\farcs28 (40~AU) along the blue jet. However, to obtain a brighter rim at the base, the emissivity must be strongly enhanced in the hemisphere facing the star. After convolution by our PSF, the map and centroid velocities are then undistinguishable from those predicted for that single hemisphere. For simplicity, we thus consider a half-sphere in the following. The predicted map is compared with the observed one in the bottom panel of Fig.~\ref{fig:cone-map}. The emission surface of (70 AU)$^2$ is close to that of the cone model.
 
For each geometry that reasonably reproduces the H$_2$ 1-0 S(1) image, we then construct a synthetic PV diagram along the jet axis, after convolving the datacube spatially by a Moffat function (a good fit to the raw PSF delivered by our AO system)  and spectrally by a gaussian of FWHM 88 km/s (median resolution of SINFONI over the field of view). We then compute velocity centroids and line widths as a function of distance in the same way as for our Figure~\ref{fig:pvh2}, i.e., by performing a gaussian fit to each PV spectrum. We find that the broad wings of the Moffat function severely smear out velocity gradients. {\em Spatial convolution by the PSF is thus essential for a meaningful comparison with observed centroids}. In contrast, spectral convolution affects only the line widths. For simplicity, velocity vectors are assumed to have a constant modulus $V_0$ and to make a constant angle $\alpha$ from the local normal to the cavity. The flow velocity parallel and perpendicular to the cavity wall are given by $V_\parallel \equiv V_0 \sin\alpha$ and $V_\perp \equiv V_0 \cos\alpha$, counted positively away from the star.

\begin{figure}
\includegraphics[width=0.5\textwidth]{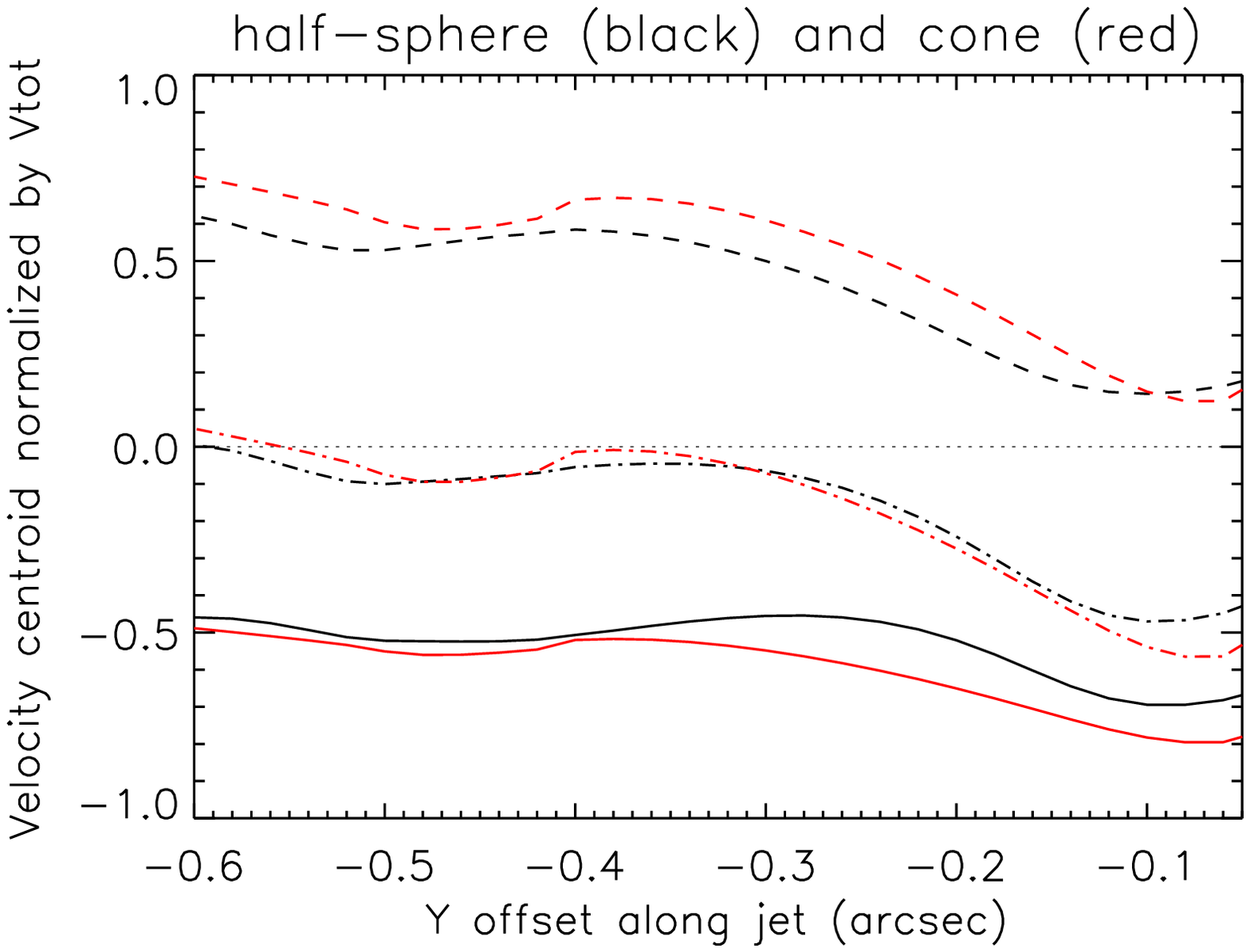}\\
\includegraphics[width=0.5\textwidth]{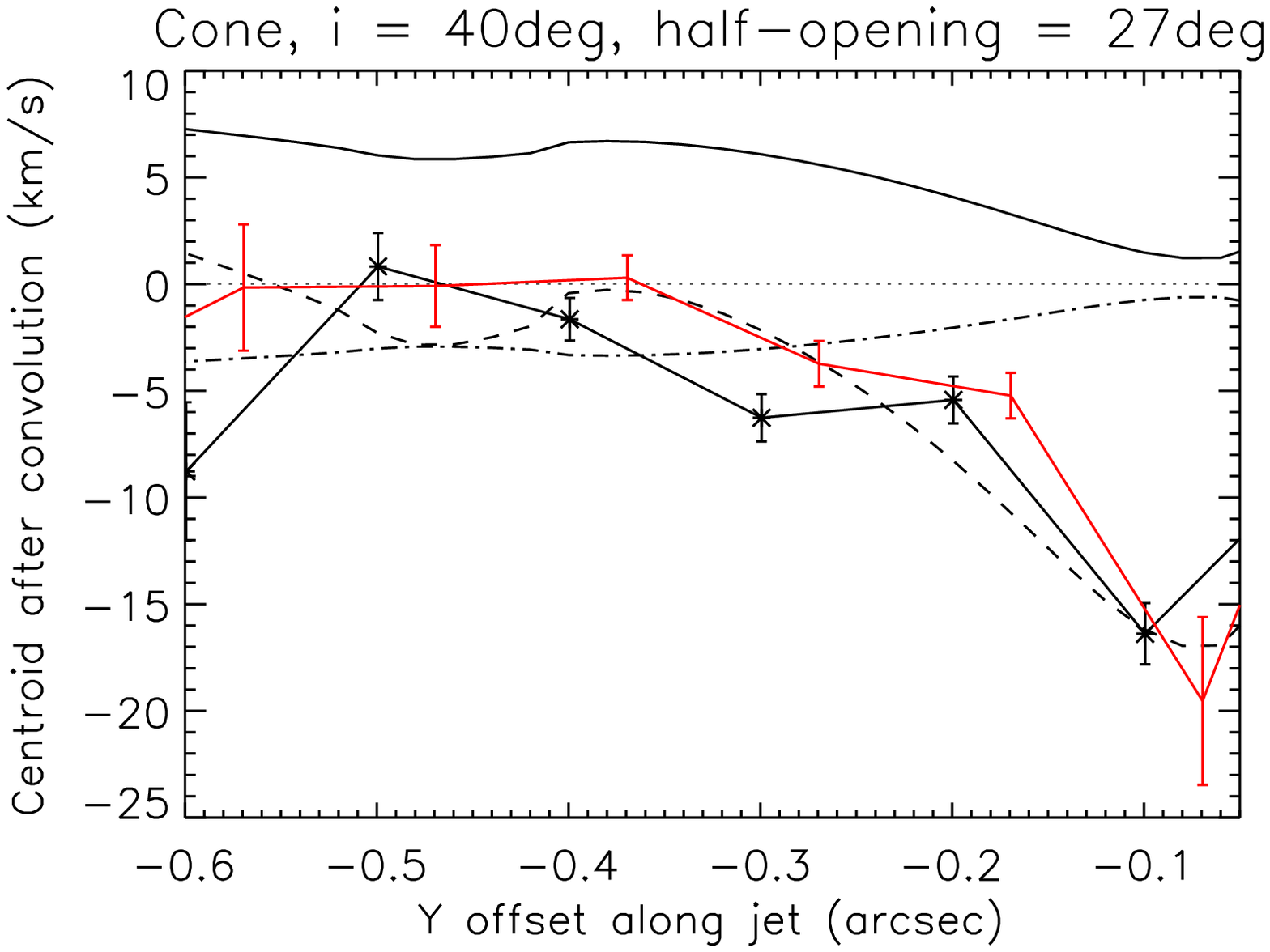}
\caption{Top panel: Predicted centroid velocities in units of the total gas speed for a PV along the jet, for the same two models shown in Fig. 8:  cone (in red) and half-sphere (in black).. The bottom solid curves are for a pure outward parallel flow ($\cos\alpha = 0$), top curves for a pure outward perpendicular flow ($\cos\alpha = 1$), and middle curves for an intermediate situation ($V_\parallel/V_\perp = 1.3$, $\cos\alpha = 0.6$). Bottom panel: Observed H$_2$ 1-0 S(1) centroid velocities in a PV along the jet axis (symbols with error bars, from Fig.~\ref{fig:pvh2}) compared with predictions for the conical model. {\it Dashed curve:} $V_\perp$ = 20 \kms\ and $\cos\alpha$ = 0.6; {\it Dash-dot curve:}. $V_\perp = -5$ \kms\ and $V_\parallel = 0$; {\it Solid curve:}  $V_\perp = +10$ \kms\ and $V_\parallel = 0$. }
\label{fig:pvcent-alpha}
\end{figure}

The predicted PV centroid velocities in units of $V_0$ are plotted in the top panel of Figure~\ref{fig:pvcent-alpha} as a function of distance along the jet, for both the cone and the half-sphere models presented in Fig.~\ref{fig:cone-map}. The results are strikingly similar despite the different geometries. This gives us confidence that the derived conclusions are robust and not dependent on the exact cavity shape. For a pure parallel flow oriented away from the source ($\cos\alpha = 0$), the PV centroid  is blueshifted at all distances by $\simeq -0.5 V_\parallel$. In contrast, for a pure perpendicular flow oriented outward ($\cos\alpha = 1$), the centroid is everywhere redshifted by an amount that increases with distance up to $\simeq +0.65 V_\perp$. 
Centroids for intermediate values of $\cos\alpha$ are simply given by a linear combination of these two extremes.
A particularly interesting case is $\cos\alpha = 0.6$ (i.e., $V_\parallel /V_\perp= 1.3$), shown by the middle curves in Figure~\ref{fig:pvcent-alpha}. This produces radial velocities that drop to essentially zero at -0\farcs4 from the star, as observed. 

In the bottom panel of Figure~\ref{fig:pvcent-alpha}, 
the observed H$_2$ 1-0 S(1) centroids in a PV along the jet (from Fig.~\ref{fig:pvh2}) are compared with those predicted for a conical surface (similar conclusions would be reached for the half-sphere). Only an intermediate value of $\cos\alpha \simeq 0.6$ (i.e., $V_\parallel /V_\perp= 1.3$) can reproduce at the same time the large blueshift observed at -0\farcs1, and the centroids close to zero observed at -0\farcs4. In this case, the data are well fitted with $V_\perp \simeq 20$\kms. A larger $V_\perp \simeq 30$\kms\ would predict excessive blueshifts close to the source. Examination of the predicted line widths yields the same result, namely that $V_\perp \simeq 20$\kms\ is compatible with the data while $V_\perp \simeq 30$\kms\ predicts excessive line broadening. The maximum flow speed compatible with the observed radial velocities and line widths is thus $V_0 = V_\perp/\cos\alpha \simeq$ 30 \kms. 

If the -20 \kms\ blueshift at -0\farcs1 is instead due to an axial jet knot, the value of $\alpha$ in the wide-angled rims is no longer well determined. The only constraint now is that PV centroids at -0\farcs2 to -0\farcs4 fall in the range -5 to 0 \kms\ (within the uncertainties discussed in Section 2). This approximately translates into the condition $(V_\parallel - V_\perp) \simeq 5$ \kms. For example, a pure {\em inward} perpendicular flow with $V_\perp \simeq -5$ \kms\ (dash-dot curve in the bottom panel of Figure~\ref{fig:pvcent-alpha}) or a pure parallel flow with $V_\parallel \simeq 5$ \kms\ (not shown) become compatible with observations within the uncertainties. 

In summary, the observed radial velocities of H$_2$ 1-0 S(1) in the PV diagram limit the gas motion perpendicular to the cavity walls  to $-5 \ge V_\perp \le +5$ \kms, unless there is a significant parallel component $V_\parallel \simeq 1.3 V_\perp$ in which case $V_\perp$ can be up to  20\kms\ and the total flow speed can reach 30\kms. The above constraints help to narrow-down the possible wind and ambient properties in various scenarios, which we now examine.



\subsection{Irradiated disc atmosphere}

\begin{figure*}
\centering
\begin{tabular}{|c|c|}
\hline
\includegraphics[angle=90, trim= 3.5cm 7cm 5.5cm 2cm,clip,width=0.47\textwidth]{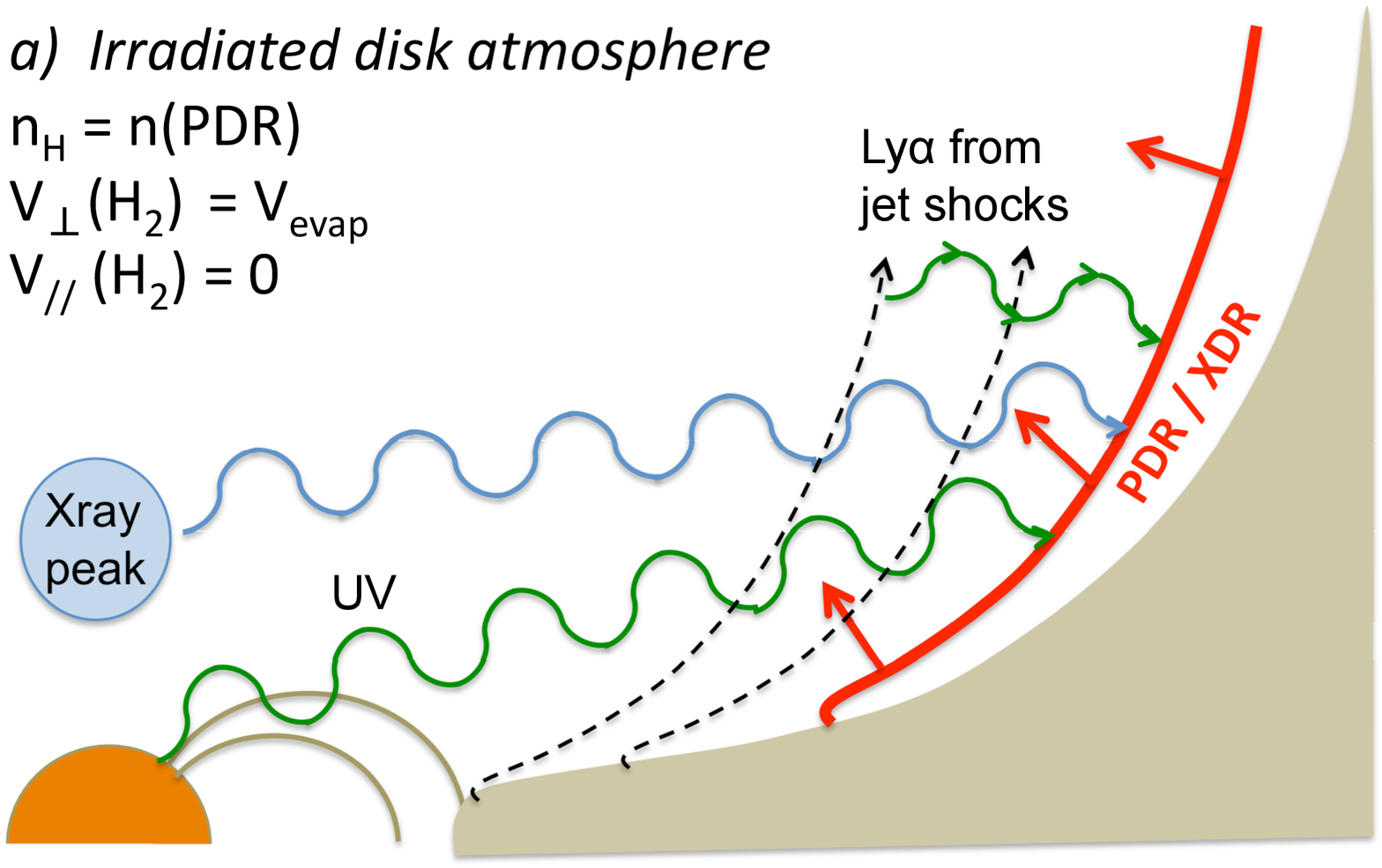} &
\includegraphics[angle=90, trim= 3.5cm 7cm 5.5cm 2cm,clip,width=0.47\textwidth]{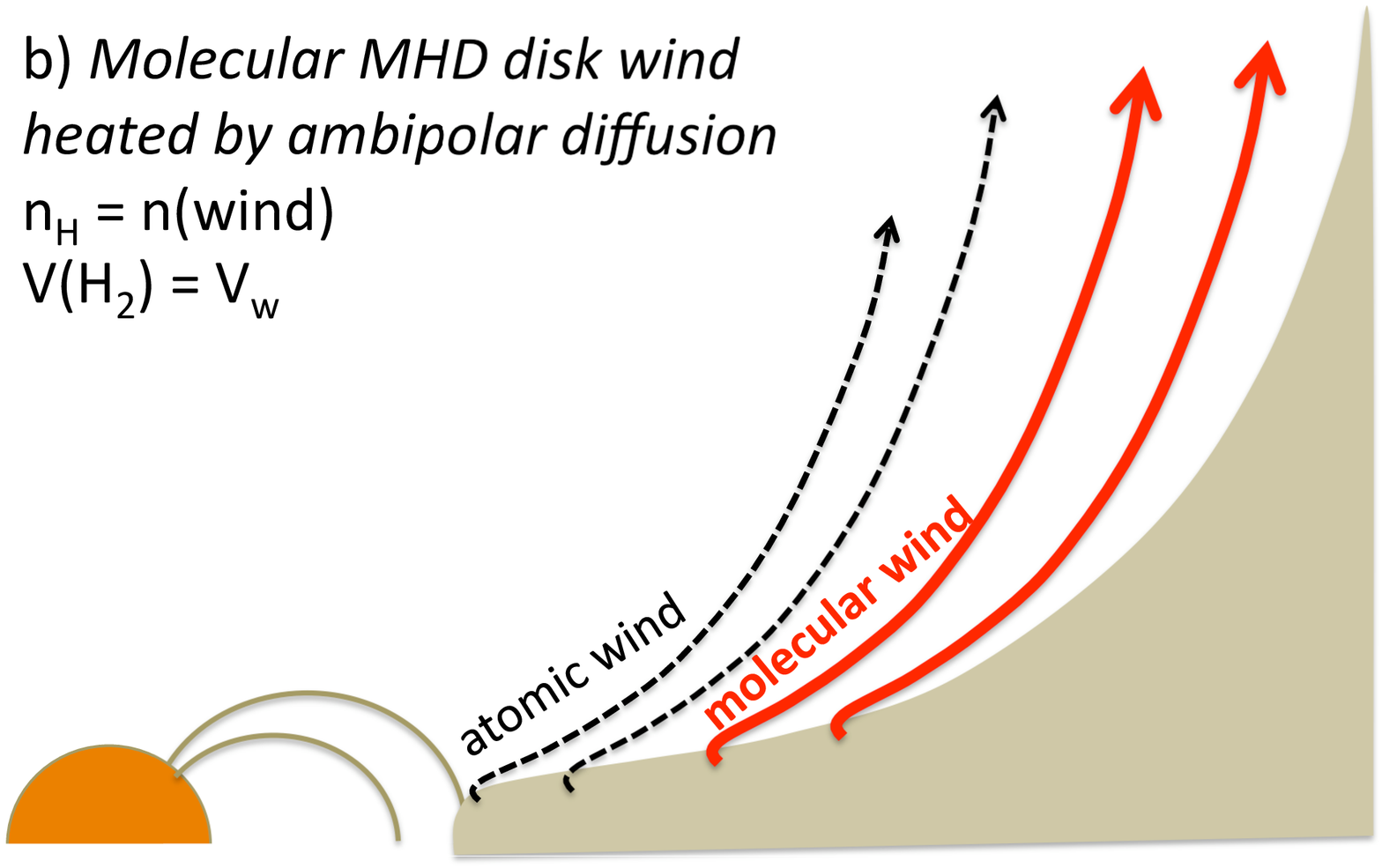} \\
\hline
\includegraphics[angle=90, trim= 3.5cm 7cm 5.5cm 2cm,clip,width=0.47\textwidth]{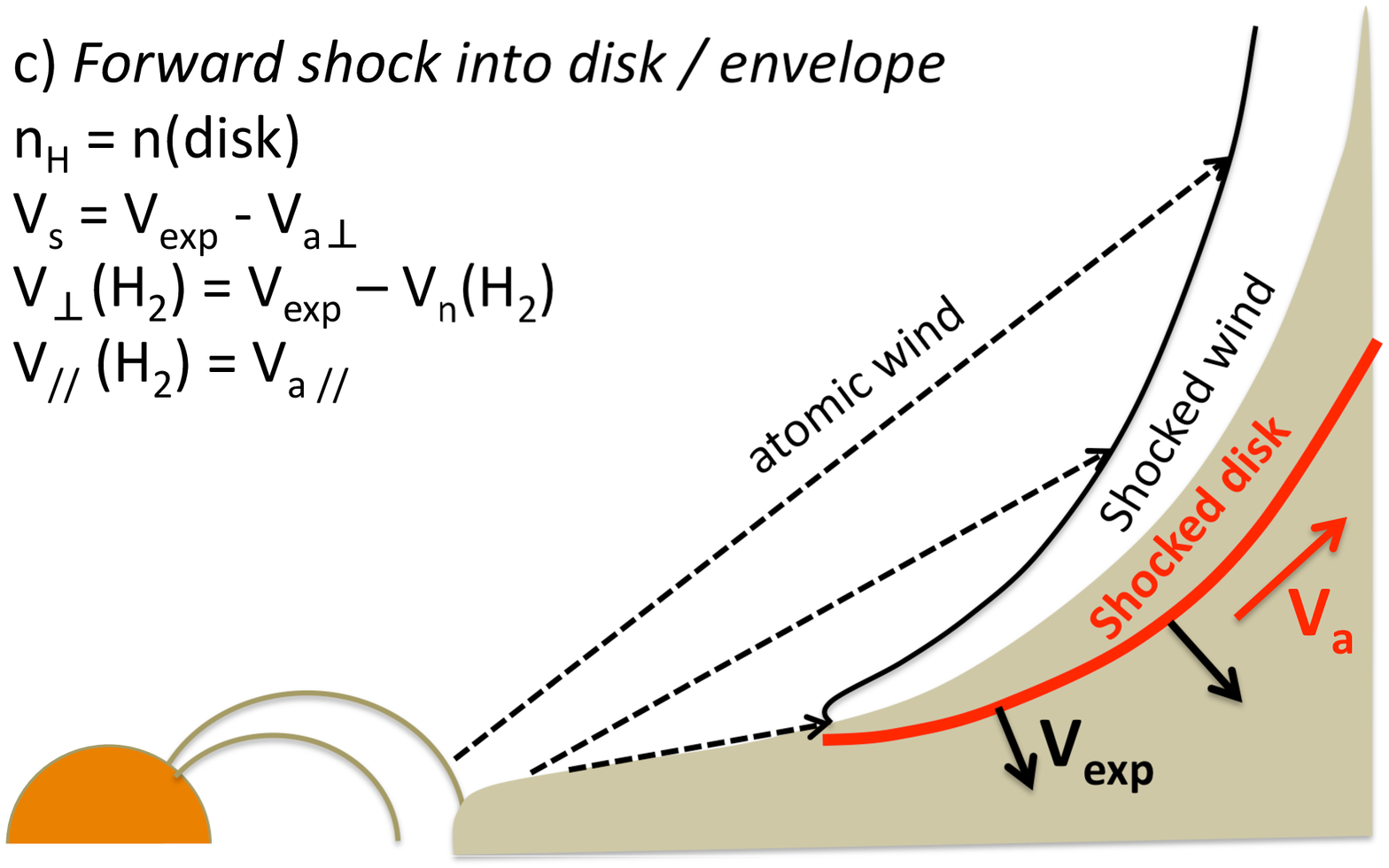} &
\includegraphics[angle=90, trim= 3.5cm 7cm 5.5cm 2cm,clip,width=0.47\textwidth]{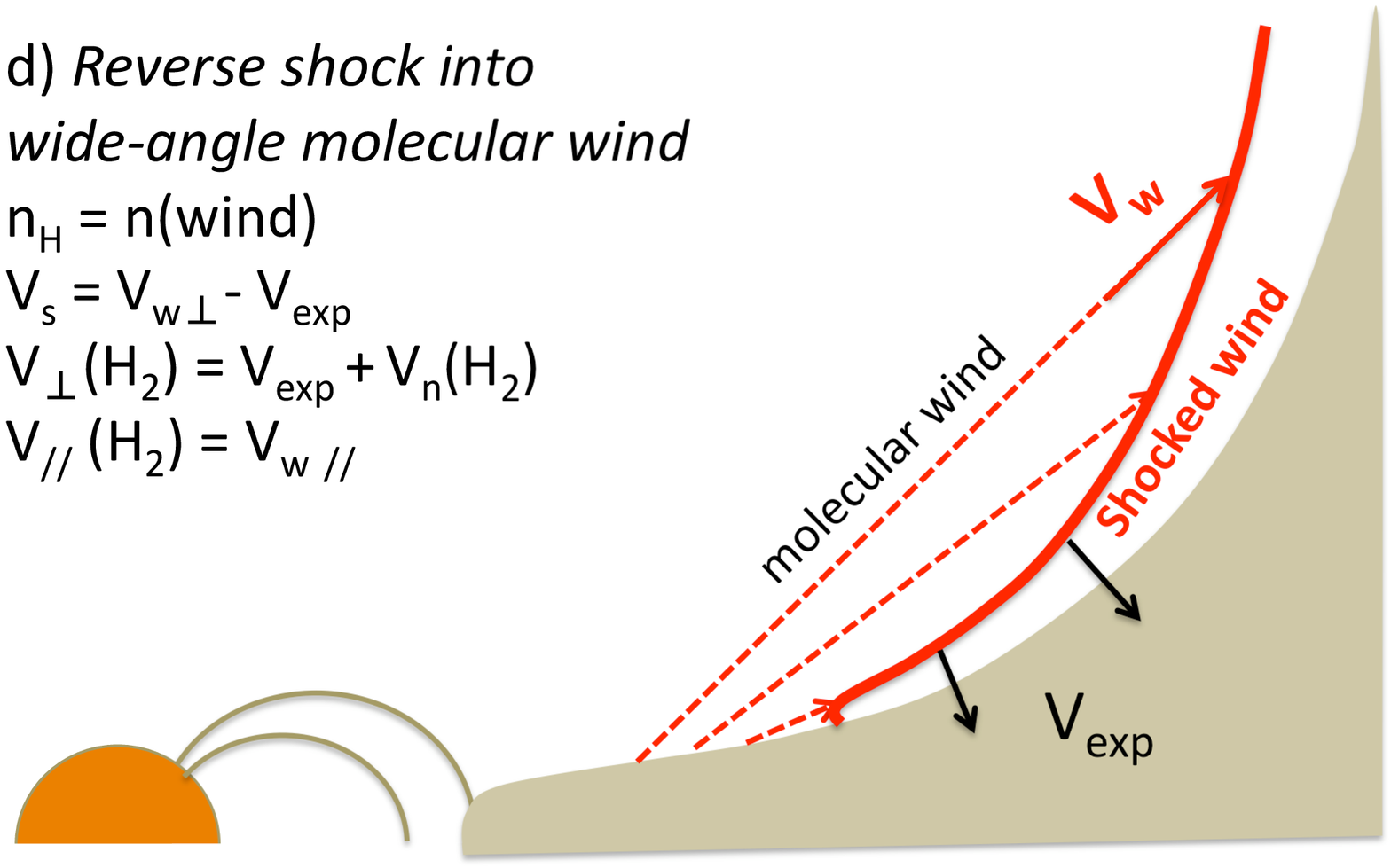} \\
\hline
\end{tabular}
\caption{Schematic representation of the 4 scenarii discussed for the origin of the wide-angle rovibrational H$_2$ emission in DG Tau. The layer of emitting H$_2$ at 2000 K is drawn in solid red. Wide-angle wind streamlines are shown as dashed (black if atomic, red if molecular). The velocity of the emitting gas perpendicular and parallel to the cavity wall ($V_\perp$ and $V_\parallel$), and the shock speed \Vs\ and preshock density \nH\ when applicable, are also listed as a function of the expansion proper motion of the cavity $V_{\rm exp}$,  the velocity of the H$_2$ 1-0 S(1) emitting layer in the frame of the shock wave, $V_n({\rm H_2})$, and the initial velocity of the ambient gas ($V_a$) or molecular wind ($V_w$).}
\label{fig:scenario_molecular}
\end{figure*}

We first reconsider the scenario depicted in panel a) of Fig.~\ref{fig:scenario_molecular}, where rovibrational H$_2$ emission in DG Tau traces a disc atmosphere heated by stellar FUV and Xray radiation. Indeed, a recent indication that the warm H$_2$ emitting in 1-0 S(1) is strongly irradiated is the presence of Ly$\alpha$-pumped H$_2$ FUV emission with the same 2D emission morphology as in 1-0 S(1) \citep{Schneider2013b} and with relative intensities among pumping levels indicating a similar gas temperature $\simeq$ 2500~K \citep{Herczeg2006}.
 
\citet{Beck2008} previously argued against stellar FUV-Xray photons as the dominant heating source for the extended ($\ge 100$ AU) H$_2$ emission around CTTS, including DG Tau, based on the irradiated disc models of \citet{Nomura2005} and \citet{Nomura2007} which predict surface temperatures less than 2000 K beyond disc radii of 20 AU. 
However, we note that the FUV stellar flux in DG Tau is higher than adopted in these models. 
The extinction-corrected  FUV excess in DG Tau is flat over at 1400-2000\AA, with a flux measured on earth of 2$\times$10$^{-13}$ \mbox{erg s$^{-1}$ cm$^{-2}${\small \AA}$^{-1}$}~\citep{Gullbring2000}. 
This corresponds to a normalized FUV flux\footnote{defined as the average ratio over 910-2066 \AA\ of the incident FUV field to the standard interstellar field  \citep[see e.g.,][]{Habing}.} $G_{0} = 5\times10^{5}$ at a distance of 0\farcs1 = 14 AU from the star,  5 times more than in the model of \citet{Nomura2007}. With this G$_{0}$ value, the PDR models of \citet{LePetit2006} predict H$_2$ line ratios compatible with observations for \nH = $10^6-10^{7}$ \cm, although the predicted 1-0 S(1) brightness for a face-on view is 4--14 times lower than observed. Towards the center of the H$_2$ cavity at 0\farcs25= 35AU, the unattenuated FUV field is $G_{0}=8 \times10^{4}$ and the predicted brightness for \nH = $10^{6}-10^{7}$ \cm\ is now only 2--5 times lower than observed, although the predicted 2-1 S(1)/1-0S(1) ratio $\simeq 0.15-0.2$ is slightly too high. Given that PDR models apply only to infinite uniform slabs, and necessarily involve some uncertainties, this may still be considered as promising agreement. Recently, \citet{Schneider2013b} argued that FUV pumping of H$_2$ in the region 0\farcs2--0\farcs4 from DG Tau may be enhanced by Ly$\alpha$ photons from the 30-100\kms\ atomic shocks known to be present in the DG Tau jet. If the Ly$\alpha$ pumping flux that they infer is uniformly spread over the emission area $\simeq 2 \times 10^{30}$ cm$^2$ estimated here (see Sect.~\ref{sec:h2column}), then $G_{0} \simeq 10^6$. This enhanced level of irradiation could reproduce {\it both} the face-on cavity brightness in 1-0 S(1) and the 2-1 S(1)/1-0 S(1) ratio if \nH = $10^{7}$ \cm\ \citep{LePetit2006}, without the need for shock excitation. 


Assuming that the warm wide-angle H$_2$  in DG Tau traces the 2000 K layer in the irradiated disc atmosphere, a photoevaporative flow would seem a natural hypothesis to explain the observed slight blueshifts. Photo-evaporation occurs mostly outside of a disc radius $r_{\rm cr} \simeq 0.15 r_{\rm g}$, where $r_{\rm g}$ is the radius where the sound speed equals the keplerian speed \citep{Dullemond2007}. For the mass of DG Tau $\simeq 0.7 M_{\sun}$ and an assumed temperature  of 2000 K at the flow base, we obtain $r_{\rm cr} = 10(\mu/2)$~AU with $\mu$ the mean molecular mass; A higher initial temperature will give a smaller $r_{\rm cr}$. This is consistent with the maximum radius $\simeq 0\farcs1 = 14$ AU at the base of the H$_2$ rims in the FUV image of \citet{Schneider2013b}. Hydrodynamical models \citep[e.g.,][]{Font2004} predict that the expected evaporation speed is on the order of a few times the sound speed $c_s = 3 (T/2000{\rm K})$ \kms\ for molecular gas. The dot-dashed curve in Fig.\ref{fig:pvcent-alpha} shows the predicted PV centroids for a photoevaporative flow directed {\em into} the cavity at 1.5 times the sound speed ($V_\perp = -5$ \kms); the predicted low blueshifts are roughly consistent with observed centroids beyond 0\farcs2 from the source; the high blueshifts of -20\kms\ closer to the star would then trace a faster axial jet knot of different origin. 

\citet{Takami2004} ruled out FUV-Xray irradiation as the excitation mechanism of the warm blueshifted H$_{2}$ in DG Tau, because they inferred a momentum flux of $6.5\times 10^{-6}$ \Msun\kms\yr\ too large compared to typical values in CO outflows from low-luminosity Class I protostars. The large momentum flux of \citet{Takami2004} stems from an assumed mean flow speed $\simeq$ 20\kms, length along the flow $\simeq$ 40 AU, emitting area of $4 \times 10^{30}$ cm$^2$, and a large total column density $N_H = 10^{21}$ cm$^{-2}$ (the gas at 2000 K in equilibrium PDRs has a low molecular abundance $\simeq 10^{-3}-10^{-2}$ \citet{Burton1990}). 
On the other hand, with our twice smaller characteristic emitting area, and longer spatial extent of 60 AU along the jet axis, the momentum flux would become only $2\times 10^{-6}$ \Msun\kms\yr, comparable to CO outflows in Class I sources of 1\Lsun\ \citep{Bontemps1996}.  
In addition, most of the warm H$_2$ at wide angle could be photoevaporating at low speed $\leq 5$ \kms\ (see above), giving an even lower momentum flux by a factor 4. Since DG Tau is a "flat spectrum" TTS at the border between Class I and Class II, this would not be prohibitive anymore. Furthermore, we caution that momentum fluxes in Class I outflows are based on CO observations in low-J lines with a large beam ($\ge 10\arcsec$ typically) and a bias towards cold gas at $T\le 100$~K. They may not be sensitive to the warm and compact outflows at 2000 K on $< 100$ AU scales probed by H$_2$ imaging. Rovibrational CO lines would be a better tracer of such conditions, and recent CO $v=1-0$ observations at high resolution do provide evidence for warm, compact, and slow molecular winds towards several high-accretion Class II sources, attributed to disk photoevaporation \citep{Bast2011}. The corresponding momentum fluxes are not known sufficiently well yet to compare with DG Tau.

We conclude that FUV irradiation does not seem fully excluded as the heating and perhaps acceleration mechanism of the warm H$_2$ wide-angle cavity in DG Tau. However, an important open issue with this scenario is the high PDR density of \nH = $10^{7}$ \cm\ needed to reproduce the observed brightness and line ratios. At radii of 10--30 AU, static models of the irradiated disk in DG Tau reach this density only at polar angles $\theta \simeq 60\degr$ \citep{Podio2013}, whereas toy models reproducing the observed morphology favor a layer much higher up from the midplane, at $\theta \simeq 30-45\degr$ (see Section~\ref{sec:toymodels}). Lower densities might be possible if the PDR is not at equilibrium, e.g., if the FUV flux was only recently "turned-on" \citep{Hollenbach1995} or if photoevaporation constantly exposes fresh H$_2$ to the FUV flux. However, the observed limit on the rim thickness $\le 0\farcs1$ combined with the typical gas column at 2000~K in PDRs of $N_H \simeq10^{21}$ cm$^{-2}$ still points to a high \nH $\ge 5 \times 10^6$ \cm. Time-dependent hydro-chemical models of disk PDRs including photoevaporation and irradiation by Ly$\alpha$ photons from jet shocks would be in order to definitely test this scenario. 

\subsection{Molecular MHD disc wind heated by ambipolar diffusion}

Based on the pioneering work of \citet{Safier1993}, \citet{Takami2004} was first to suggest that the wide-angle slow H$_2$ emission in DG Tau might trace the outer molecular streamlines of an MHD disc wind heated by ambipolar diffusion, whose inner atomic streamlines would produce the faster and more collimated emission seen in optical lines. This scenario is depicted in panel b) of Fig.~\ref{fig:scenario_molecular}. 

 \citet{Panoglou2012} recently carried out a thorough study of the thermal and chemical structure of a steady, self-similar MHD disc wind solution from \citet{Casse2000}, with a lever arm parameter $\lambda = 13$ selected to provide a good fit to the tentative rotation signatures in the atomic jet of DG Tau \citep{Pesenti2004}. 
For an accretion rate $\dot{M}_{acc} \sim 10^{-6}-10^{-7}$ \Msun\yr, Panoglou et al. find that H$_{2}$ molecules 
can survive against both collisional dissociation and photodissociation by the stellar FUV and Xray photons when launched from disc radii greater than about 1 AU. Heating by ambipolar diffusion then creates a temperature plateau of 2000-4000 K over 10--100 AU scales, in good agreement with the observed uniform H$_2$ temperature in DG Tau (see Sect.~\ref{sec:h2column}).  For a launch radius $r_0 \simeq 10$~AU, compatible with the rim radius at the base of the blue lobe, this particular MHD disc wind solution predicts that the streamline will reach a radius $\simeq$ 50 AU at an altitude of 50 AU above the disc (see Fig.~1 of \citet{Panoglou2012}), similar to the measured maximum FWHM of the H$_2$ cavity at the corresponding projected distance (see Fig.~\ref{fig:fwhm}). The predicted rotation speed is $V_\phi \simeq 5$ \kms, compatible with our upper limit on transverse rotation signatures (see Sect.~\ref{sec:vshift}). At the same position, the streamline reaches a poloidal flow velocity $V_0 \sim$ 15 \kms~ consistent with the constraint $V_0 \leq 30$ \kms\ from modeling of radial velocities (see Sect.~\ref{sec:toymodels}).

Assuming that the wide-angle H$_2$ emission does trace a fully molecular MHD disk wind uniformly heated to 2000 K by ambipolar diffusion, the one-sided wind mass flux may be calculated following the method of \citet{Takami2004}, assuming LTE. Our toy models favor a slightly larger spatial scale along the jet axis $R_{max} \simeq$ 60 AU, giving ${\dot M}_w \sim 10^{-9}$ \Msun\yr. With an estimated accretion rate of ${\dot M}_{\rm acc} = (3\pm 2)\times 10^{-7}$ \Msun\yr\ over the period 1996--2005 \citep{Agra-Amboage2011}, the ejection-accretion ratio in the blue lobe of the molecular wind would thus be $\simeq$ 0.003--0.014. This would be consistent with the lever arm parameter $\lambda = 13$ of the MHD disc wind solution if the external launch radius, $r_{\rm e}$, and internal launch radius, $r_{\rm i}$, for the molecular streamlines are in a ratio $r_{\rm e}/r_{\rm i} \simeq$ 1.15-2 (see, e.g., Eq. (17) of \citet{Ferreira2006}). The corresponding width of the flow would be 50\%-13\% of the cavity radius, compatible with observations. We infer $r_{\rm i} \simeq$ 5--8~AU for $r_{\rm e}\simeq$ 10 AU. This value of $r_{\rm i}$ matches rather well with the outer launch radius $\simeq$ 3 AU for the atomic LVC component inferred from rotation signatures \citep{Pesenti2004}. 

Hence, the temperature, kinematics, morphology, and mass flux of the H$_2$ cavity in DG Tau all appear in promising agreement with a steady molecular MHD disc wind heated by ambipolar diffusion, ejected from 5--10 AU. However, calculations of beam-convolved synthetic H$_2$ images and PV diagrams of MHD disc winds taking into account non-LTE effects, the full 3D velocity field, and a range of streamlines are necessary to test this scenario. Such detailed work is beyond the scope of the present paper and will be the subject of future work.




\subsection{Forward shock driven into the environment}

In this scenario, depicted in panel c) of Fig.~\ref{fig:scenario_molecular}, the observed wide-angle H$_2$ emission traces ambient gas from the surrounding disc  / envelope, that is, being shocked and swept-up into a shell by the wide-angle atomic wind. This scenario is traditionally invoked to explain CO molecular outflow cavities seen on larger scales around young protostars \citep{Arce2007}. It was proposed by \citet{Takami2007} to explain the small scale V-shaped biconical cavity of H$_2$ 1-0 S(1) emission around HL Tau, which bears strong resemblance to that in DG Tau; it is thus interesting to investigate whether the same scenario could apply to DG Tau itself. In the following discussion, we will neglect turbulent mixing along the cavity walls between the shocked ambient gas and  the shocked wind. Current prescriptions for the entrainment efficiency, and numerical simulations of wind-envelope interaction, both suggest that it would not be efficient on the scales of interest \citep{Delamarter2000}. 

 We place ourselves in the reference frame of the star, where we denote as $V_{\rm exp}$ the expansion proper motion of the shock wave into the envelope, and $V_{\rm a}$ the initial velocity of the ambient gas counted positively away from the star. We may then write the shock speed $V_s$ suffered by ambient gas, and the velocity of postshock warm H$_2$ perpendicular and parallel to the shock front, as 
\begin{eqnarray}
V_s & = & V_{\rm exp} - V_{\rm a\perp} \\
V_\perp  &= &V_{\rm exp} - V_n(H_2) = V_s + V_{\rm a\perp} - V_n(H_2) \\
V_\parallel & = & V_{\rm a\parallel},
\end{eqnarray}
where $V_n(H_2) \ge 0$ is the centroid velocity of the shock-heated 1-0 S(1) emission layer {\em in the reference frame of the shock wave} (see Sect.~\ref{sec:shocks}). 

Let us first consider the simplest case of a static ambient medium ($V_{\rm a} =0$), where the shock speed $V_s = V_{\rm exp}$. Only a J-shock driven at $V_s=$10\kms\ into a dense disk/envelope with $n_H \simeq 10^{6}$ \cm\ and $b \leq 0.1$ can match at the same time the constraint from proper motions that $V_{\rm exp} < 12$ \kms (see Sect.~\ref{sec:map}) and the observed H$_2$ rovibrational excitation and brightness (see Table~1). Therefore, the cavity would have expanded to its observed half-width of 0\farcs2 = 30 AU in only $\simeq$ 15 years, meaning that it would trace a very recent wind ``outburst", even younger than in the case of HL Tau \citep[70 years;][]{Takami2007}. Since it would be improbable to catch a rare event at such an early age, this phenomenon would have to be recurrent on timescales of a few decades, similar to the successive ``bubbles" seen in XZ Tau \citep{Krist2008}. 

However, a strong caveat with a shocked static preshock is that it predicts redshifted H$_2$ velocity centroids: In a J-shock, $V_n(H_2) \simeq 0$, hence the shock-heated H$_2$ is moving perpendicular to the shock surface with $V_\perp = V_s \simeq 10$ \kms\ while $V_\parallel = 0$. The predicted centroids are plotted as a solid curve in the bottom panel of Fig.~\ref{fig:pvcent-alpha} for a conical toy model (similar results are obtained for a half-sphere). They are redshifted by $\simeq +4$ to +6 \kms\ at distances of -0\farcs2 to -0\farcs4 from the star, whereas observed centroids at the same distances are $\simeq$ -5 to 0\kms. The discrepancy is in the same sense in both exposures, and larger than our maximum possible residual errors in wavelength calibration and correction for uneven slit illumination (see Section 2). Therefore, a forward shock into a static ambient medium does not seem able to explain the observed centroids satisfactorily. 

Let us now consider the case where ambient gas surrounding the cavity would be infalling towards DG Tau at speed $V_{\rm ff}$ (about 5 \kms\ at a mean distance of 45 AU (0\farcs3) for a 0.7$M_\odot$ star). The V-shape of the cavity walls means that the infalling stream will have a very oblique incidence, with a dominant component parallel to the shock and directed towards the star. The terms $V_{\rm a\perp} = -V_{\rm ff \perp}$ and $V_{\rm a\parallel} = -V_{\rm ff \parallel}$ are then both negative, with the latter largely dominant. The effect on postshock centroids is to add a component with negative $V_0 = -V_{\rm ff}$ and $\cos\alpha < 0.6$ ($V_{\rm a\parallel}/V_{\rm a\perp} > 1.3$). Fig.~\ref{fig:pvcent-alpha} shows that this will produce additional redshift, worsening the discrepancy with observed centroids compared to the static ambient case. Therefore this scenario can be ruled out.


Let us now consider the opposite case where the ambient medium would be initially outflowing, and $V_{\rm a\perp}$ and $V_{\rm a\parallel}$ are both positive. To keep a shock speed $V_s \simeq 10$ \kms\ as required for H$_2$ excitation, the outflow perpendicular to the wall should be very slow at $V_{a\perp} = V_{\rm exp} - V_s < 2$ \kms\ (see Eq.~1). Therefore $V_\perp$ would remain essentially unchanged from the static case, at 10--12 \kms\ (see Eq.~2). But redshifted centroids could  be avoided if the (conserved) outflow parallel to the shock front $V_{a\parallel} = V_{\parallel}$ is  about 1.3$V_\perp \simeq 13-15$ \kms\ (see curve for $\cos\alpha = 0.6$ in Fig.~\ref{fig:pvcent-alpha}, scaled to $V_0 \simeq 17-20$\kms). The mass flux in this preshock "ambient" outflow would be: $\mu m_H n_H V_{a\parallel} 2\pi r \Delta r$ $= 2 \times 10^{-8} (2\Delta r/r) (r/30\, {\rm AU})^2 M_\odot$ yr$^{-1}$. The origin of such a pre-existing massive molecular flow is an open issue. Previous ejecta from older outburst episodes would be expected to move mainly perpendicular to the shock surface, rather than parallel as needed to match the centroids. A photoevaporative flow from further out in the disk also seems difficult, as the column density of shocked gas swept-up over 15 yr, $N_H = n_H V_s t = 5 \times 10^{20}$ cm$^{-2}$,  would absorb a significant fraction of FUV photons before they reach the outer disk surface. A slow MHD-driven wind from the outer disk would remain an option; it is noteworthy that submm observations of CO and C$^{+}$ towards DG Tau do show signs of slow expansion with a high-velocity tail reaching up to -10 \kms, albeit on scales of 5--10\arcsec\ much larger than those probed here \citep{Kitamura1996, Podio2013}.

In summary, the detailed spatial, velocity, and brightness information extracted from our SINFONI data shows that  a forward shock sweeping up ambient gas may explain the wide-angle  H$_2$ 1-0 S(1) emission in DG Tau, but only under a set of restrictive circumstances:  it should trace a very young (15 yr) swept-up shell expanding at $V_{\rm exp} \simeq$10-12 \kms\ into a pre-existing dense ($\simeq 10^6$\cm) molecular wind moving mainly along the cavity walls at $\simeq$ 13-15 \kms. The origin of this preshock ambient outflow is an open issue. Accurate proper motion measurements of $V_{\rm exp}$ would provide a first crucial test of this scenario, and could also reveal deceleration indicative of an outburst origin, as well as possible recurrence.

\subsection{Reverse shock into a wide-angle molecular wind}

 \citet{Beck2008} and \citet{Schneider2013b} proposed that the V-shaped H$_2$ emission in the blue lobe of DG Tau traces a "shocked molecular wide-angle wind" surrounding the atomic LVC. In panel d) of Fig.~\ref{fig:scenario_molecular}, we illustrate the case where  the "reverse" shock is driven into the molecular wind as it encounters the disc / envelope. Such a geometry is suggested by the smooth curved morphology of the emission (instead of a series of knots). The forward shock driven in the environment is assumed slow enough, or of low enough density, to have negligible emission in H$_2$ 1-0 S(1).  
The shock speed $V_s$ and the postshock velocity of warm H$_2$ perpendicular and parallel to the shock front (again in the reference frame of the star) are now given as a function of the wind speed, $V_{\rm w}$, by 
\begin{eqnarray}
V_s & = & V_{\rm w\perp} - V_{\rm exp} \\
V_\perp  &= &V_{\rm exp} + V_n(H_2) =  V_s + V_{\rm w\perp} + V_n(H_2)  \\
V_\parallel & = & V_{\rm w\parallel}.
\end{eqnarray}
 Here, the constraint that $V_\perp \le 20$\kms\ set by observed line centroids and line widths (see Sect.~\ref{sec:toymodels}) imposes the same limit on $V_n(H_2)$. This is only fulfilled by J-shocks (where $V_n(H_2) \simeq 0$) and by the C-shock with $b=0.5$ in Table~1 (where $V_n(H_2) \simeq 0.6V_s = 18$ \kms).  The latter is excluded, however, as it predicts a well resolved emission thickness of 64 AU = 0\farcs45 (FWHM) in H$_2$ 1-0 S(1), while images and toy models indicate a barely resolved rim $\le 0\farcs1$. Note that this does not exclude a strongly magnetized wind with $b > 0.1$. All C-shock models in Table~1 assume a preshock ionization from cosmic rays alone, as appropriate for ambient gas. But a molecular wind launched within 10 AU is irradiated by stellar FUV photons and Xrays and will reach a higher ionization $\simeq 10^{-4}$ that remains "frozen-in" as the wind expands \citep[e.g.,][]{Glassgold1991,Panoglou2012}. A J-shock will then result even if $b > 0.1$, due to the strong ion-neutral coupling. The required J-shock speed and/or preshock density would just increase with $b$, to compensate for the reduced post-shock compression (higher magnetic pressure).

 A definite advantage of the "reverse shock" scenario, compared to the "forward shock" discussed previously, is that the required shock speed can be reached even with a static cavity ($V_{\rm exp} =0$) by adjusting $V_{\rm w\perp}$. This would allow a lower proper motion than the current upper limit of 12 \kms, an an older cavity age than the 15 yr implied in the forward shock case.  A second advantage is that the (conserved) wind component parallel to the shock, $V_{\rm w\parallel}$, naturally tends to produce blueshifted H$_2$ centroid velocities; these will be compatible with observations if $V_{w\parallel} - V_\perp \simeq 5$ \kms (see Sect.~\ref{sec:toymodels}).

Despite its advantages, a serious caveat with the reverse-shock scenario is that the required wind mass flux seems excessive. Since the shock emitting in H$_2$ 1-0 S(1) is the wind-shock, the one-sided mass flux \Mw\ intercepted by the shock surface is directly given by  
\begin{eqnarray}
\dot{M}_{\rm w} &=&  (\mu m_{\rm H} A_{\rm s}) n_{\rm H} V_{\rm w\perp}\\
&=& 6.7 \times 10^{-8} \frac{n_{\rm H}}{10^6 {\rm cm}^{-3}} \frac{V_{\rm w\perp}}{10  {\rm km\,s}^{-1}} M_\odot {\rm yr}^{-1},
\label{eq:mdot-wind-shock}
\end{eqnarray}
where we took again a shock area $A_s \simeq 2\times 10^{30}$ cm$^2$. The scalings are for the minimum possible values of  \nH\ and $V_{\rm w\perp}$, corresponding to the case $b=0.1$ and $V_{\rm exp} =0$. This lower limit is already uncomfortably close to the DG Tau accretion rate, given that the result should be doubled to account for the occulted red lobe of the wind. A higher $b$ in the wind would require a higher preshock density and/or shock speed and thus an even larger \Mw. The value of $V_{\rm w\perp}$ would also double if the cavity expands at $V_{\rm exp} = 10$ \kms\ instead of being static. 

We conclude that the "reverse shock" scenario for the origin of the wide-angle H$_2$ v=1-0 emission in DG Tau has several attractive features compared to a forward shock (possibility of a static cavity, naturally blueshifted centroids). However, it appears to require an excessive wind mass flux compared to current accretion rates in DG Tau. 

\section{Conclusions}\label{sec:conclusions}

We have conducted a thorough analysis of H$_{2}$ 1-0 S(1) spectro-imaging data towards the classical T Tauri star DG Tau, with an angular resolution of 0\farcs12 and a precision in velocity centroids reaching down to a few \kms. We compared the morphology and kinematics with simultaneous data in [Fe II] \citep{Agra-Amboage2011}, near-contemporaneous spectra in [O I] \citep[][]{Coffey2007}, and spectra and images in FUV-pumped H$_2$ obtained 6 years later \citep{Schneider2013a, Schneider2013b}. The absolute line brightness and the 2-1 S(1) / 1-0 S(1) ratio were used to constrain shock conditions, based on state-of-the-art shock models \citep{Kristensen2008}. Projection effects in radial velocities were modeled to constrain the intrinsic velocity field of the emitting gas. This new information was used to re-examine proposed origins for the wide-angle H$_2$ emission in the blue lobe of DG Tau and similar sources. Our main conclusions, and our suggestions for further work, are the following: 
\begin{itemize}

\item The faint arc of H$_2$ 1-0 S(1) emission in the red lobe traces slow material at +5 \kms\ in the wings of a large bowshock prominent in [Fe~II]. In contrast, the limb-brightened H$_2$ 1-0 S(1) at the base of the blue lobe is much wider than the [Fe II] jet; it appears filled-in by the wide-angle atomic wind at low velocities seen in [O I] and [S II], that is, steady over $\ge 4$yr but invisible in [Fe II] probably because of strong iron depletion. Most of the wide-angle H$_{2}$ 1-0 S(1)  is at low blueshifts (between -5 and 0 \kms). Hence, we confirm previous claims that the H$_2$ blue lobe in DG Tau appears as a  slower, wider extension of the nested velocity structure previously observed in atomic lines \citep{Takami2004, Schneider2013b}. The H$_2$ may be rotating about the jet axis in the same sense as the disk and atomic flow, with $V_\phi \le 7$ \kms.

\item The blue lobe morphology in H$_{2}$ 1-0 S(1) is strikingly similar to the FUV image \citep{Schneider2013b}, confirming that they trace the same hot gas at 2000~K and setting upper limits on the rim thickness $\le 14$~AU and on any expansion proper motion $V_{\rm exp} < 12$\kms. The face-on surface brightness in 1-0 S(1) is $\simeq 3 \times10^{-3}$ erg s$^{-1}$cm$^{-2}$ sr$^{-1}$, and the 2-1 S(1) / 1-0 S(1) ratio is uniform at our resolution. If due to shock-heating, they require a J-shock at 10 \kms\ into $10^6$ \cm\ gas (C-shocks would predict excessive rim thickness). Modeling of the radial velocities limits the gas motion perpendicular to the walls  to $-5 \le V_\perp \le +5$ \kms, unless there is a comparable parallel component in which case $V_\perp$ can reach  20\kms. 
  
\item A photoevaporating disk atmosphere irradiated by the strong stellar FUV excess and Ly$\alpha$ from jet shocks seems able to explain the excitation and brightness on observed scales, as well as the low blueshifts over most of the area. One caveat is the high density in the upper disk atmosphere of \nH = $10^{7}$ \cm\  suggested by steady PDR models. Time-dependent PDR models including disc photoevaporation may alleviate this problem and would be highly desirable. 

\item A molecular MHD disc wind launched around 10 AU and heated by ambipolar diffusion is another promising scenario that can reproduce the observed morphology, kinematics (including limits on rotation), mass flux rate, and temperature  \citep[see][]{Panoglou2012}. Synthetic predictions in H$_2$ 1-0 S(1) taking into account non-LTE effects, the full 3D velocity field, and a range of streamlines are necessary to fully test this hypothesis and will be the subject of future work.

\item A cavity of shocked ambient swept-up gas does not reproduce the H$_2$ 1-0 S(1) excitation and blueshifted centroids unless it is very young (15 yr) and expands into a pre-existing dense molecular wind of unclear origin moving at $\simeq$15 \kms\ along the cavity walls. A crucial check of this scenario would be to detect the predicted proper motion $V_{\rm exp} \simeq 10$\kms, and the signs of deceleration or recurrence expected for such an episodic phenomenon. 

\item A reverse shock into a wide-angle molecular wind could readily explain the excitation, low blueshifts, and apparent stationarity of the H$_2$ cavity, but the required one-sided wind mass flux of $\ge 6.7 \times 10^{-8} M_\odot$ yr$^{-1}$ appears excessive compared to current accretion rates in DG Tau.
\end{itemize}
We point out that the (J-type) shock scenarios predicts a FWHM of only 0.1 AU for the H$_2$ 1-0 S(1) emitting layer. This is to be compared with a predicted thickness of 7 AU for a static PDR ($N_H \simeq 10^{21}$ cm$^{-2}$ and \nH $\simeq 10^7$ \cm) and 4--14 AU for an MHD disk wind. Spatially resolving the rims of the H$_2$ cavity would thus bring useful insight into its origin. We also caution that we have considered simplistic scenarios where only one heating mechanism is operating at a time. A combination of shock and FUV irradiation might alleviate the caveats encountered when each is considered separately. 






\begin{acknowledgements}
The authors are grateful to D. Coffey for kindly providing the [O I] transverse slit data shown in Fig.~\ref{fig:pv_trans_feii-h2-oi}, to G. Herczeg for precious advice on velocity calibration, and to the referee, M. Takami, for insightful comments that led to significant improvement of the paper. V. Agra-Amboage acknowledges financial support through the Funda\c c\~ao para a Ci\^encia e Tecnologia (FCT) under contract SFRH/BPD/69670/2010. This research was partially supported by FCT-Portugal through Project PTDC/CTE-AST/116561/2010. The authors also acknowledge financial and travel support through the Marie Curie Research Training Network JETSET (Jet simulations, Experiments and Theory) under contract MRTN-CT-2004-005592, and through the french Programme National de Physique Stellaire (PNPS).
\end{acknowledgements}

\bibliographystyle{aa}
\bibliography{biblio}

\appendix

\section{Correction of SINFONI data for uneven slit illumination}
\label{app:uesi}

\begin{figure}[!h]
\includegraphics[angle=90,width=0.5\textwidth]{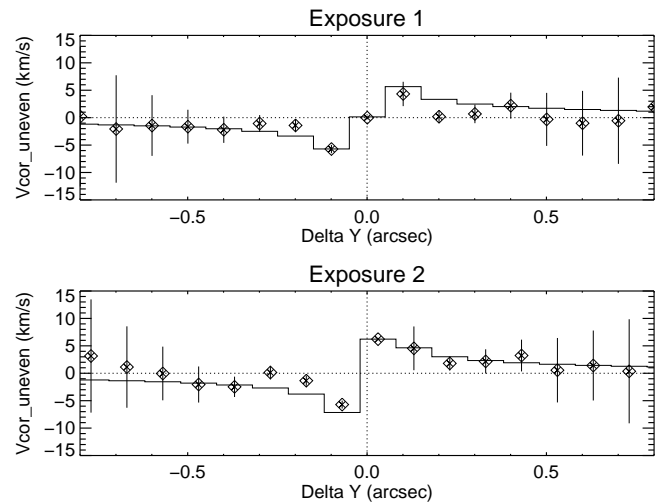}
\caption{Velocity corrections for uneven slit illumination in PV diagrams along the jet averaged $\pm 0\farcs5$ across the jet axis for exposure 1 (top) and exposure 2 (bottom). Black histograms plot the theoretical correction computed for a 2D brightness distribution fitted by a Moffat function. Symbols with error bars show empirical shifts obtained from cross-correlation of telluric absorption features against the reference spectrum (at $Y = 0\arcsec$ in exposure 1 and $Y = +0\farcs03$ in exposure 2). To ease comparison with the theoretical model, they are shifted vertically to match the model at the reference spectrum position.}
\label{fig:fit_vcor_uneven}
\end{figure}

 We obtain an empirical estimate of the uneven-slit illumination effect in our K-band data by noting that telluric absorption features will be subject to the same spurious shifts as the H$_2$ lines, since the light distribution in the slitlet is dominated by the stellar continuum in both cases.  To estimate the shifts in telluric lines as a function of distance from the continuum peak, we increase signal to noise by constructing a position-velocity (PV) diagram along the jet, averaging spaxels over a width $\pm 0\farcs5$ across the jet axis (i.e., along each slicing mirror). We then cross-correlate each of these PV spectra with the spectrum closest to the DG Tau position, over an interval of 0.06$\mu$m left of the H$_2$ line that shows numerous telluric absorption features. The resulting empirical shifts in each exposure is plotted as diamond symbols in Fig.~\ref{fig:fit_vcor_uneven}. It may be seen that the effect is $\simeq \pm 5$\kms\ for the slitlets within $\pm 0\farcs1$ of the source and drops to $\le$ 2\kms\ beyond this region. We also modeled the uneven-slit illumination effect for each exposure following the formulation developed by \cite{Marconi2003}, using a Moffat function fitted to the corresponding continuum image to represent the 2D light distribution delivered by the AO. The modeled velocity correction is overplotted as black histograms in Fig.~\ref{fig:fit_vcor_uneven}.  The agreement with the velocity shifts obtained by cross-correlation is seen to be very good, with a maximum deviation $\simeq$ 2.0 \kms\ within $\pm 0\farcs3$ of the star and compatible with zero further out.  We therefore use this 2D illumination model to correct for uneven slit illumination the H$_2$ velocity centroids in each individual spaxel (Fig.~\ref{fig:centroid_maps}), in the PV diagram along the jet (Fig.~\ref{fig:pvh2}), and in a transverse PV-diagram across the jet (Fig.~\ref{fig:pv_trans_feii-h2-oi}). We expect residual systematic errors in the corrected velocities of at most $2$\kms\ within $\pm 0\farcs3$ of the star and less than this further out, where the effect anyway becomes quite small. 



\section{Shocks models}\label{an:shocks_models}

\begin{figure}
\includegraphics[width=0.5\textwidth]{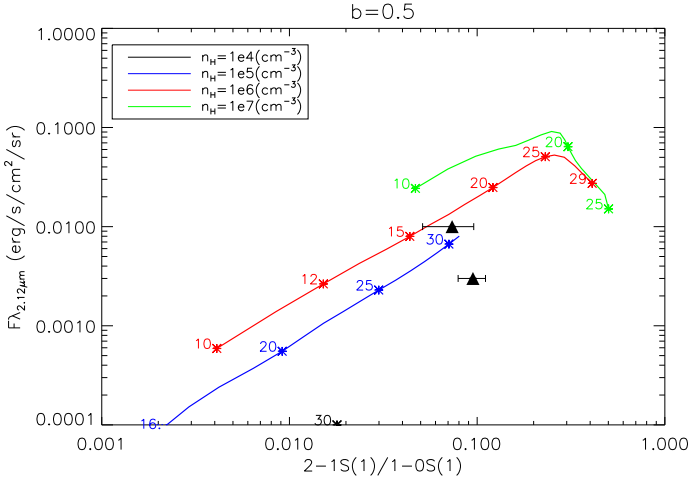}

\caption{Plot of the H$_{2}$ brightness in the $v=$ 1-0 S(1) 2.12 $\mu$m line against the $v=$2-1 S(1)/$v=$1-0 S(1) ratio for two positions in the DG Tau blue lobe (filled triangles with error bars) and for the grid of
planar C-shock models with $b$ = 0.5 (color curves) calculated by \citet{Kristensen2008}. Each curve corresponds to a different value of the preshock hydrogen nucleus density, increasing from bottom to top: \nH = $10^{4}$, $10^{5}$, $10^{6}$ and $10^{7}$ cm$^{-3}$. The shock speed \Vs\ increases to the right and some values are marked along the curves to guide the eye.}
\label{fig:c-shocks_modelb05}
\end{figure}

\begin{figure}
\centering
\includegraphics[width=0.5\textwidth]{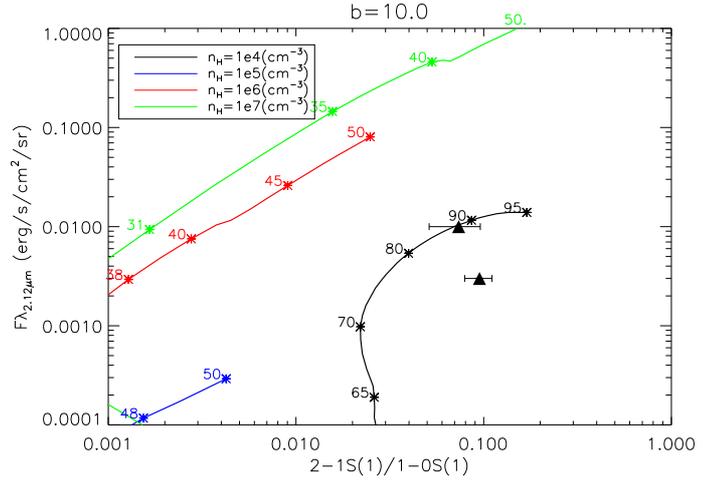}
\caption{Same as Fig.~\ref{fig:c-shocks_modelb05} but for a grid of C-shocks with $b$=10.0.}
\label{fig:c-shocks_modelb10}
\end{figure}

The grid of shock models computed by  \citet{Kristensen2008} provides the expected H$_2$ line fluxes for different input values of \nH, \Vs, and magnetic field parameter $b$. Comparing the observed $v= 1-0$ S(1) brightness and $v=2-1$ S(1)/ $v=1-0$ S(1) ratio with those predicted, the preshock density and shock velocity can be derived, depending on the magnetic field. These constraints are used in Section~\ref{sec:discussion} to discuss scenarios involving shocks for the origin of the  H$_2$ emission in DG Tau. We presented in Fig.~\ref{fig:molec_model} the model predictions for J-shocks with $b$=0.1 and  C-shocks with $b$=1, as an example. Here we present in Fig.~\ref{fig:c-shocks_modelb05} and Fig.~\ref{fig:c-shocks_modelb10} similar model predictions for $b$=0.5 and $b$=10, two extreme cases that  give a more complete view of the behavior of these models and explain the results shown in Table~\ref{tab:shocks}. 

\end{document}